\documentclass[11pt]{article}

\usepackage{bm}
\usepackage[normalem]{ulem}
\usepackage{amsmath,graphics,epsfig,float,braket,multirow,url}
\usepackage{caption}
\usepackage{subcaption}
\usepackage{comment}
\usepackage{authblk}
\usepackage[capitalize]{cleveref}
\usepackage[normalem]{ulem}

\newcommand{\s}[1]{{\textsf{\textbf{#1}}}}

\begin{document}

%%%% Article title to be placed here
\title{\s{Compliant Mechanisms for Invertible Poisson’s Ratio and Tunable Stiffness  in Cell Culture Substrates}}
\author{ \textsf{Manu Sebastian $^\dagger$, Sreenath Balakrishnan$^{*}$, and Safvan Palathingal $^\dagger$}}
\date{{\it *School of Mechanical Sciences,
    Indian Institute of Technology, Goa,
    India}\\
{\it $^{\dagger}$Department of Mechanical and Aerospace Engineering,\\ Indian Institute of Technology Hyderabad, Telangana, India}\\[2ex]
\today
}

 \maketitle
\hrule\vskip 6pt

%%%% Abstract text to be placed here %%%%%%%%%%%%
\begin{abstract}
The mechanical environment of a substrate plays a key role in influencing the behavior of adherent biological cells. Traditional tunable substrates have limitations as their mechanical properties cannot be dynamically altered in-situ during cell culture. We present an alternate approach by using compliant mechanisms that enable realization of tunable substrate properties, specifically, invertible Poisson's ratio and tunable stiffness. These mechanisms transition between positive and negative Poisson's effects with tunable magnitude through a bistable Engaging-Disengaging Compliant Mechanism (EDCM). EDCM allows stiffness between two points of the substrate to switch between zero and theoretically infinite. In the stiffened state, lateral deformation reverses under a constant axial load, while in the zero-stiffness state, the deformation direction remains outward as that of re-entrant structure. EDCM in conjunction with an offset mechanism also allows tuning of the effective stiffness of the entire mechanism. We present analytical models correlating geometric parameters to displacement ratios in both bistable states and through illustrative design cases, demonstrate their potential for designing dynamic and reconfigurable cell culture substrates.  
\end{abstract}

\vskip 6pt
\hrule
\vskip 6pt
\section{Introduction}

Biological cells are influenced by the mechanical properties of their environment. When stem cells were cultured on substrates with varying elastic moduli, those on softer substrates differentiated into brain-like cells, while those on stiffer substrates transformed into bone-like cells, reflecting the elastic properties of these tissues—soft for the brain and rigid for bone \cite{Engler2006}. This seminal finding sparked widespread interest in exploring how cell behavior relates to the elastic properties of substrates. Initial experiments involved creating soft substrates (with elastic moduli ranging from 1 to 100 kPa) using polyacrylamide gels \cite{Engler2006}, hydrogels \cite{DRURY20034337}, or PDMS pillars \cite{Saez2007}, then culturing cells on them and analyzing differences in cell responses. While these studies provided valuable insights, they overlooked the dynamic remodeling of the extracellular matrix driven by factors like aging, inflammation, fibrosis, and diseases such as cancer.

To address this dynamic environment, new methods were developed to investigate how cells respond to changes in the extracellular matrix. In one approach, cells were detached (via trypsinization) from substrates with high elastic modulus and transferred to those with lower elastic modulus \cite{Yang2014}. However, this detachment and reattachment process complicated result interpretation. To resolve these issues, photoactivatable hydrogels were introduced, allowing the elastic modulus to be adjusted through irradiation while cells remained attached \cite{Kloxin2009}. Compared to the extensive research on the effect of elastic modulus of cell substrate on cell function, fewer studies have explored the effect of Poisson’s ratio \cite{Yan2017,Song2018,Zhang2013}. This gap likely stems from two factors: (i) cells primarily respond to mechanical properties at low elastic moduli (1–100 kPa), and (ii) the typical material design techniques for low-modulus substrates, such as gels and pillars, do not easily allow Poisson’s ratio to be manipulated. In studies \cite{Yan2017,Song2018}, polyurethane scaffolds with negative Poisson’s ratios were shown to promote vascular and neuronal differentiation in pluripotent stem cells. In \cite{Zhang2013}, a re-entrant structure crafted via two-photon polymerization demonstrated that negative Poisson’s ratio can disrupt normal cell division. Yet, to our knowledge, no techniques exist for modifying a substrate’s Poisson’s ratio during cell growth. Additional drawbacks of current materials-based approaches for substrate modification include: (i) the challenge of repeatedly switching elastic moduli (e.g., high-low-high-low cycles), and (ii) the limited range of elastic properties that can be toggled. 

To address these challenges, we have devised a novel approach for adjusting the elastic properties of substrates during cell growth, utilizing compliant micromechanisms equipped with bistable switches. Cells anchor to various points in their microenvironment via protein complexes called focal adhesion \cite{Geiger2009}. By exerting pushing and pulling forces at these specific attachment sites, cells assess the elastic characteristics of their surroundings \cite{Plotnikov2012}. Consequently, cells respond to the force-displacement dynamics at these focal adhesions rather than directly to the substrate’s inherent elastic properties, such as Young’s modulus or Poisson’s ratio. This insight allows us to replicate desired substrate elasticity by engineering a mechanism between these attachment points to produce a specific force-displacement relationship. 

\begin{figure*}[h]
    \centering
    \includegraphics[width=\linewidth]{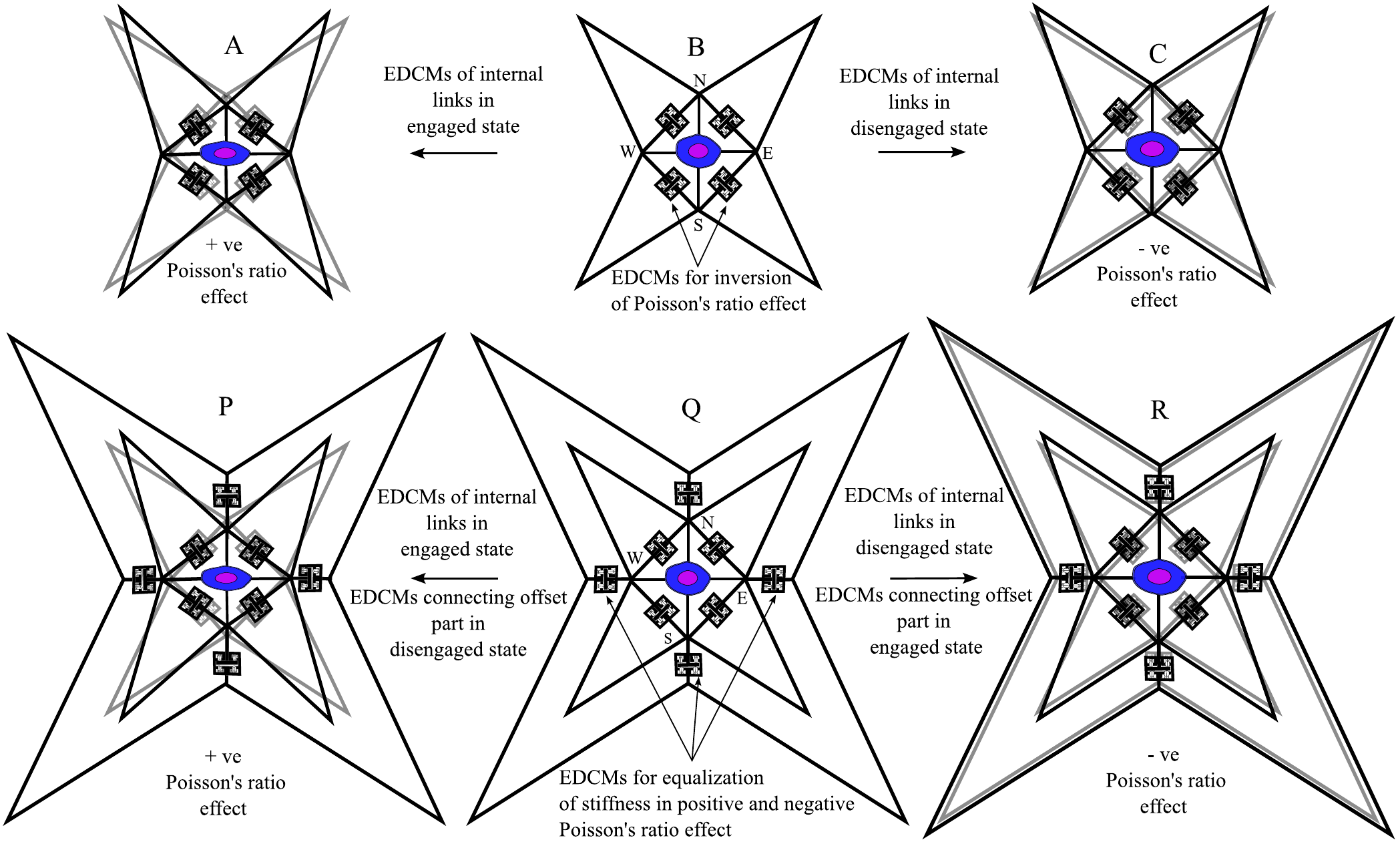}
    \caption{Compliant mechanism with invertible Poisson’s ratio. (A-C) Internal EDCMs toggle the connection between re-entrant points to switch Poisson’s behavior. (P-R) Offset mechanism with EDCMs enables stiffness tuning by engaging or disengaging offset links.}
    \label{fig:illustration}
\end{figure*}

In this work, we present the design of  a compliant mechanism-based substrate to culture cells, which is capable of switching between positive and negative Poisson’s ratios.  Building on our prior design of a single-cell biaxial stretcher based on re-entrant geometries \cite{fartyal2021,Marwah2023,Marwah2024}, we noted that when cells exert an outward force horizontally, the vertical direction also expands outward, giving the impression of a substrate with a negative Poisson’s ratio (Fig. \ref{fig:illustration}C). In contrast, when the re-entrant points are connected, the mechanism behaves as if it has a positive Poisson’s ratio (see \cref{fig:illustration}A). Motivated by this observation, we first developed a preliminary mechanism design capable of inverting the Poisson's ratio effect using slender beam elements \cite{manu2023}. In the present work, we extend that design by demonstrating how this inversion can be actively switched and the resulting stiffness tuned by integrating bistable switches--Engaging Disengaging Compliant Mechanisms (EDCM) \cite{mehul2023}-- between the re-entrant points (N, S, E, and W) of the mechanism, as illustrated in Figs. \ref{fig:illustration}A-C. 

A shortcoming of this design is the associated change in effective stiffness when switching between Poisson states. This variability can introduce indeterminacy in studying cell behavior, as it becomes difficult to isolate observed behavioral changes solely to changes in the Poisson's ratio effect.
To address this, we design the mechanism to have similar stiffness in both states by increasing the stiffness of the negative Poisson's state to the range of the positive state. We do this by incorporating another re-entrant shaped offset to the mechanism (\cref{fig:illustration}Q), which only gets engaged to the mechanism via EDCM during the negative Poisson's state as shown in \cref{fig:illustration}R. This modification ensures that both states retain their intended Poisson’s behavior while exhibiting comparable stiffness, denoted as $K$. Furthermore, we extend the analytical model to include non-slender beams since the micromechanism design obtained with  microfabrication constraints may not always exhibit high slenderness ratios.

 In \cref{sec:modelling}, we present the analytical model for the mechanisms based on Castigliano’s first theorem, modeling the links of the mechanism as beam elements  and incorporating bending, axial, and shear strain energy contributions. This approach ensures that the design is valid for a wide range of beam geometries and slenderness ratios.  In \cref{sec:Design},  we detail the design methodology through two illustrative examples: one focusing solely on switching the Poisson’s ratio effect, and the other considering both Poisson’s ratio inversion and stiffness equalization. In \cref{sec:Validation}, we validate the designs through both finite element analysis (FEA), and tabletop experiments conducted on 3D-printed prototypes.
\begin{figure*}[!htb]
    \centering
    \includegraphics[width=\linewidth]{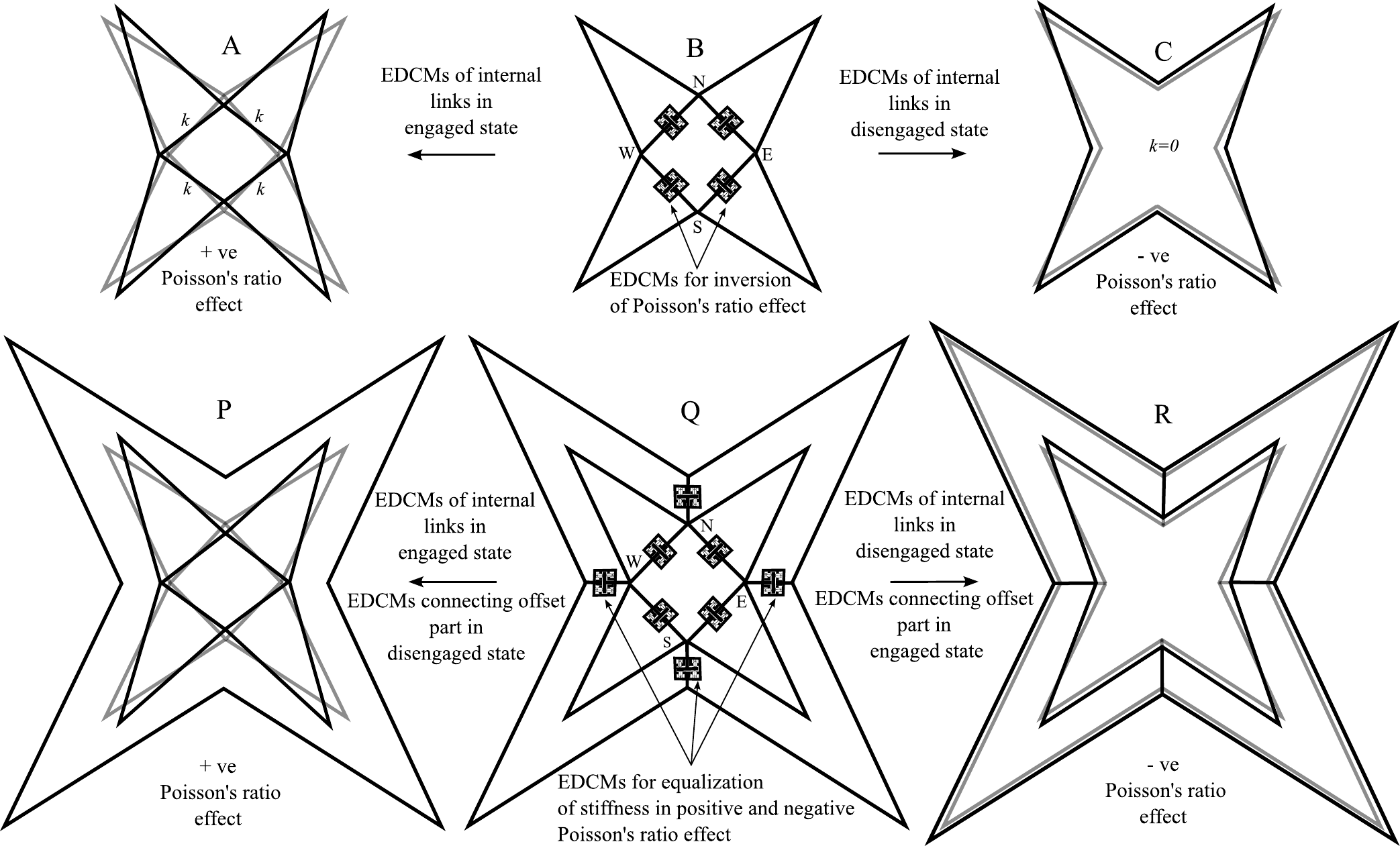}
\caption{(A-C) Beam models of the re-entrant mechanism: links with engaged EDCMs are modeled as rigid (high stiffness), while disengaged EDCMs are represented as absent (zero stiffness). (P-R) Beam models of the offset mechanism with EDCMs in engaged and disengaged states.}
    \label{fig:MechanismDesign}
\end{figure*}
\Cref{sec:EDCM} presents a third example, showcasing a fully integrated prototype with EDCMs, and compares its analytical, experimental, and FEA results. We consider a final example in \cref{sec:BiologicalAspect} and discuss fabrication constraints relevant to microfabrication using photolithography, particularly for cell-culture applications. We conclude in \cref{sec:SummaryAndFutureWork} with key findings and directions for future research.
\section{Analytical modeling}\label{sec:modelling}
In this section, we develop a mathematical model that helps design the geometrical parameters required to achieve a desired Poisson's ratio effect (expressed in terms of stretch ratio) and stiffness. The stretch ratio, which is conceptually the  inverse of Poisson's ratio,  is defined as the ratio of the horizontal displacement of the re-entrant points W or E to the vertical displacement of points N or S (see \cref{fig:MechanismDesign}B).
 When the vertical points (N and S) move outward in response to a horizontal stretch—i.e., when the horizontal points are pulled apart—the stretch ratio is considered positive (see \cref{fig:MechanismDesign}C). In contrast, if the vertical points move inward when the horizontal points are stretched, the stretch ratio becomes negative (see \cref{fig:MechanismDesign}A).

The re-entrant geometry (without stiffness compensation) used to develop the analytical model for stretch ratio is shown in Figs. \ref{fig:MechanismDesign}A and \ref{fig:MechanismDesign}C. The links of the mechanism are modeled as beam elements. To simplify the analysis, links with engaged EDCMs are modeled as beam elements with
axial stiffness $k$ (see \cref{fig:MechanismDesign}A). When the EDCMs are in the disengaged state, the links are considered absent—i.e., their stiffness is set to zero in the model—as illustrated in \cref{fig:MechanismDesign}C. 

As mentioned before, to equalize the effective stiffness $K$ experienced by cells adhered to the re-entrant points in both positive and negative Poisson's ratio configurations, an offset mechanism is integrated into the main mechanism, as illustrated in \cref{fig:MechanismDesign}Q. In the positive stretch ratio mode, this offset mechanism is engaged by activating the EDCMs in the connecting links, resulting in the equivalent model shown in \cref{fig:MechanismDesign}R. In contrast, during the negative stretch ratio mode, the offset mechanism is disengaged by deactivating the EDCMs, yielding the model shown in \cref{fig:MechanismDesign}P, effectively being the same as  \cref{fig:MechanismDesign}A.  

We first develop the analytical model focusing solely on the stretch ratio, then discuss the extended  model that incorporates stiffness tuning of the mechanism.
%%%%%%%%%
\subsection{Analytical modeling for stretch ratio}
\begin{figure}[!htbp]
    \centering
    \includegraphics[width=0.4\linewidth]{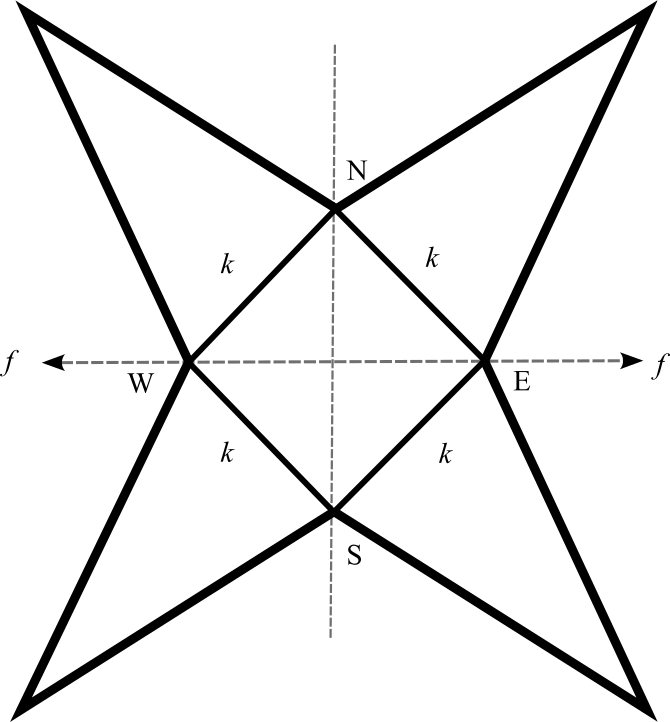}
    \caption{Compliant mechanism with beams connecting the re-entrant points.}
    \label{fig:Mechanism with links as beams}
\end{figure}

\begin{figure}[!htbp]
    \centering
    \includegraphics[width=0.4\linewidth]{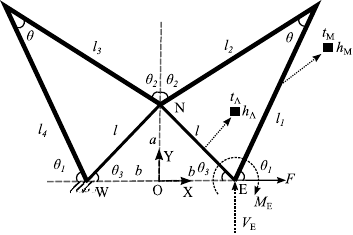}
    \caption{Geometrical parameters of the mechanism: $a, b, \theta_1, \theta_2$, $h_{\Lambda}, t_{\Lambda}$, $h_{\text{M}}, t_{\text{M}}$. In the enlarged cross-sectional view, $h_{\Lambda}$ and $h_{\text{M}}$ denote the out-of-plane dimensions, while $t_{\Lambda}$ and $t_{\text{M}}$ represent the in-plane dimensions. 
    } 
    \label{fig:Geometrical parameters}
\end{figure}

Let us consider the analytical model  given in  \cref{fig:Mechanism with links as beams} where beams connect the re-entrant points, N, E, S, and W. 
We intend to determine the displacement relation of points N and E under an applied force $f$ as shown in   \cref{fig:Mechanism with links as beams}. We utilize symmetry to simplify the model and consider only half of the mechanism, with the point W fixed, as illustrated in  \cref{fig:Geometrical parameters}.  To ensure that the half mechanism exhibits the same deflection as the full mechanism, the applied force $F$ in  (\cref{fig:Geometrical parameters}) is taken as twice the force $f$.

\begin{figure}[!htbp]
\centering\includegraphics[width=\linewidth]{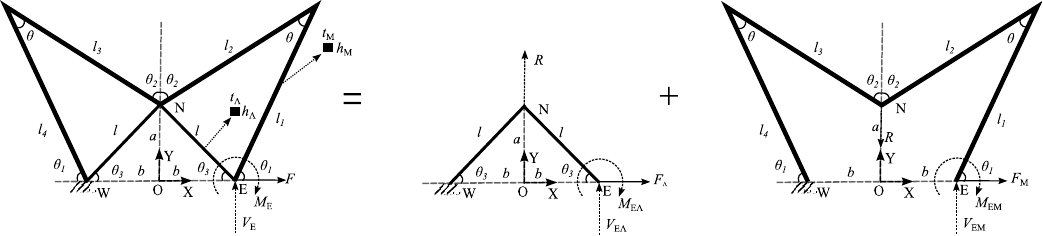}
\caption{Frame divided as the parallel combination of $\Lambda$ and M part}
\label{fig:M and V part}
\end{figure}

To further simplify this problem, we  split the mechanism into a parallel combination of two sub mechanisms: an outer M-shaped frame (referred to as the M part) and a  $\Lambda$-shaped frame  (referred to as the $\Lambda$ part),  as depicted in Fig. \ref{fig:M and V part}.
 The applied load $F$ is then distributed into two components: $F_{\Lambda}$, acting on the $\Lambda$ part, and $F_{\text{M}}$, acting on the M part, with $R$ being the reaction force at point N due to the connection between the two parts.

In the following subsections, we find the displacements of the $\Lambda$  and M parts separately using Castigliano's theorem. We then use compatibility conditions and superposition principle to derive the required force-displacement relationship for the whole mechanism.

\subsubsection{$\Lambda$ part}\label{subsectionVpart}
\begin{figure}[!htbp]
    \centering
    \includegraphics[width=0.8\linewidth]{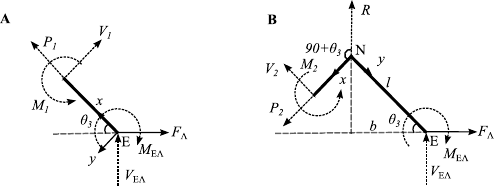}
    \caption{Sectional forces in the links of $\Lambda$ part, (A) Link1, (B) Link2}
    \label{fig:Lambda part sections}
\end{figure}

To determine the displacements of the re-entrant points E and N in the $\Lambda$ part using Castigliano's theorem, we first find the total strain energy. This is obtained by adding the strain energy contributions from stretching, bending, and shear in each link. Towards this, we find the sectional forces and moments within the individual links of the mechanism.
For the first link (Fig. \ref{fig:Lambda part sections}A),  the force and moment balance equations are given by,
\begin{align}
    F_{\Lambda}-P_{1} \cos (\theta_3)+V_{1} \sin (\theta_3)=&0,
    \label{eq:VLinkHbalance}\\
    P_{1}\sin (\theta_3)+V_{1} \cos (\theta_3)+V_{\text{E}\Lambda}=&0, ~\text{and}
    \label{eq:VLinkVbalance}\\
    M_{1}-M_{\text{E}\Lambda}+F_{\Lambda}x\sin(\theta_3)+V_{\text{E}\Lambda}x\cos(\theta_3)=&0.
    \label{eq:VLinkMbalance}
\end{align}
Solving these three equations simultaneously for $P_{1}, V_{1}$ and $M_{1}$ gives: 
\begin{align}
    P_{1}=&F_{\Lambda} \cos (\theta_3)-V_{\text{E}\Lambda} \sin (\theta_3),
    \label{eq:FC1}\\
    V_{1}=&-\cos(\theta_3) (V_{\text{E}\Lambda} + F_{\Lambda} \tan(\theta_3)),
    \label{eq:VC1}\\
    M_{1}=&M_{\text{E}\Lambda} - V_{\text{E}\Lambda} x \cos(\theta_3) - F_{\Lambda} x \sin(\theta_3)
    \label{eq:MC1}.
\end{align}
Similarly for Link 2,  the sectional forces, $P_{2}$, $V_{2}$ and moment $M_{2}$ are given by:
\begin{align}
    P_{2}=&F_{\Lambda} \cos (\theta_3)+(R+V_E) \sin (\theta_3),
    \label{eq:FC2}\\
    V_{2}=&F_{\Lambda} \sin (\theta_3)-(R+V_E) \cos (\theta_3),
    \label{eq:VC2}\\
    M_{2}=&-a F_{\Lambda}-b V_E+F_{\Lambda} x \sin (\theta_3)+ 
    M_{\text{E}\Lambda}-x (R+V_E) \cos (\theta_3).
    \label{eq:MC2}
\end{align}

The total strain energy in the links of the $\Lambda$ part, $SE_{\Lambda}$, is  given by
\begin{align}
        SE_{\Lambda}=\sum_{i=1}^{2}\left[\int\limits_0^{l_i}\left(  \frac{M_{i}^2}{2YI_{\Lambda}} 
        +\frac{\tau_i^2A_{\Lambda}}{2G}
        +\frac{\sigma_i^2A_{\Lambda}}{2Y}\right)dx_i\right]
    \label{eq:TotalSELambdapart}
\end{align}
where, $l_i$ is the length of the $i^{th}$ link, $A_{\Lambda}=t_{\Lambda}\times h_{\Lambda}$ is the cross-sectional area, and $I_{\Lambda}=\frac{h_{\Lambda}t_{\Lambda}^3}{12}$ is the second moment of area. Axial stresses, $\sigma_1$ and $\sigma_2$, are given by $\frac{P_{1}}{h_{\Lambda}t_{\Lambda}}$, and $\frac{P_{2}}{h_{\Lambda}t_{\Lambda}}$, respectively. Shear stresses, $\tau_1$ and $\tau_2$ are given by, $\tau_1=\frac{Q^2V_{1}}{I_{\Lambda}h_{\Lambda}},~\text{and}~
    \tau_2=\frac{Q^2V_{2}}{I_{\Lambda}h_{\Lambda}},$
where $Q=\frac{h_\Lambda}{2}(y^2-\frac{t_\Lambda^2}{4})$ is the first moment of area of the cross-section.

To make $SE_{\Lambda}$ given by \cref{eq:TotalSELambdapart} amenable for applying Castigliano’s theorem, we express the reactions $V_{\text{E}\Lambda}$ and $M_{\text{E}\Lambda}$ in terms of $F_{\Lambda}$ and $R$. We do this by utilizing the symmetry of the mechanism, i.e.,  the cross-section’s rotation about point E is zero. Also,  the vertical displacement at point E is zero. Therefore,
\begin{align}
    \frac{dSE_{\Lambda}}{dM_{\text{E}\Lambda}} = 0, ~\text{and}~
    \frac{dSE_{\Lambda}}{dV_{\text{E}\Lambda}} = 0.      
\end{align}
Solving these  equations helps to write $M_{\text{E}\Lambda}$ and $V_{\text{E}\Lambda}$ as:
\begin{align}
    M_{\text{E}\Lambda}=\alpha_{M_{\text{E}\Lambda}} F_{\Lambda}+\beta_{M_{\text{E}\Lambda}} R,
    \label{eq:MEVPart}\\
    V_{\text{E}\Lambda}=\alpha_{V_{\text{E}\Lambda}} F_{\Lambda}+\beta_{V_{\text{E}\Lambda}} R.
    \label{eq:VEVPart}   
\end{align}
Here $\alpha_{M_{\text{E}\Lambda}}$, $\beta_{M_{\text{E}\Lambda}}$, $\alpha_{V_{\text{E}\Lambda}}$, and $\beta_{V_{\text{E}\Lambda}}$ are coefficients that depend on geometrical parameters\footnote{These coefficients are introduced for brevity. Full expressions for these and subsequent coefficients appearing in this manuscript are provided in the Supplementary Information.}. By substituting \cref{eq:MEVPart,eq:VEVPart} in \cref{eq:TotalSELambdapart}, we get strain energy of $\Lambda$ part in terms of $F_{\Lambda}$ and $R$.

To find the displacement in the X direction, $\delta_{\Lambda_X}$, we differentiate $SE_{\Lambda}$, with respect to the applied force, $F_{\Lambda}$, 
\begin{align}
    \delta_{\Lambda_X}=\frac{\partial{SE_{\Lambda}}}{\partial{F_{\Lambda}}}
    =A_{X_{F_{\Lambda}}}F_{\Lambda}+B_{X_R}R.
    \label{eq:Dispalcement_X_LambdaPart}
\end{align}
Similarly, the displacement in the Y direction at point N, $\delta_{\Lambda_Y}$, is obtained by differentiating  $SE_{\Lambda}$ with respect to the reaction force, $R$, i.e.,
\begin{align}
     \delta_{\Lambda_Y}=\frac{\partial{SE_{\Lambda}}}{\partial{R}}
    =A_{Y_{F_{\Lambda}}}F_{\Lambda}+B_{Y_R}R.
    \label{eq:Dispalcement_Y_LambdaPart}   
\end{align}
$A_{X_{F_{\Lambda}}}$, $B_{X_R}$, $A_{Y_{F_{\Lambda}}}$, and $B_{Y_R}$   depend on the geometrical parameters of the mechanism. 

\subsubsection{M part}\label{subsubsection:MPart}
\begin{figure*}
        \centering\includegraphics[width=1\linewidth]{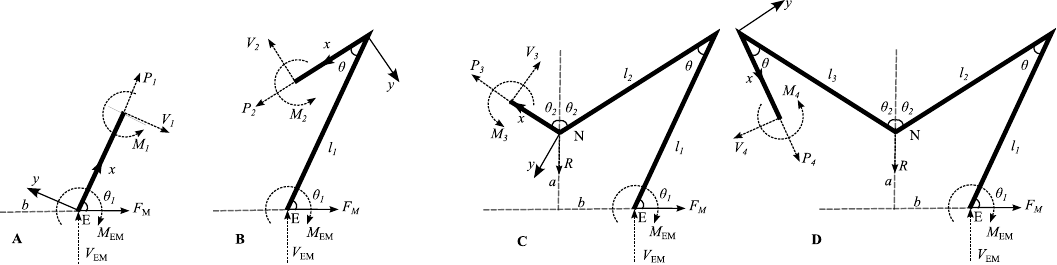}
    \caption{Sectional forces and moments in the links of M part, (A) link1, (B) link2, (C) link3, (D) link4}
    \label{fig:M part Sections}
\end{figure*}
 We obtain sections forces and moment in terms of the support reactions $V_{\text{EM}}$, $M_{\text{EM}}$ and applied force $F_{\text{M}}$ for the all the links of the M part from force balance (see \cref{fig:M part Sections}) given by: 
\begin{align}
    P_{1}=&-(V_{\text{EM}}+F_{\text{M}}\cot(\theta_1))\sin(\theta_1),
    \label{eq:MpartFC1}\\
    V_{1}=& V_{\text{EM}}\cos(\theta_1) - F_{\text{M}} \sin(\theta_1),
    \label{eq:MpartVC1}\\    
    M_{1}=&M_{\text{EM}} + V_{\text{EM}} x \cos(\theta_3) - F_{\text{M}} x \sin(\theta_1),
    \label{eq:MpartMC1}\\
     P_{2}=&F_{\text{M}}\cos(\theta-\theta_1)-V_{\text{EM}}\sin(\theta-\theta_1)
    \label{eq:MpartFC2},\\
    V_{2}=&-F_{\text{M}}\sin(\theta-\theta_1)-V_{\text{EM}}\cos(\theta-\theta_1)
    \label{eq:MpartVC2},\\
    M_{2}=&M_{\text{EM}}-V_{\text{EM}}x\cos(\theta-\theta_1)+L_1V_{\text{EM}}\cos(\theta_1)
    -F_{\text{M}}x\sin(\theta-\theta_1)-F_{\text{M}}L_1\sin(\theta_1)   
    \label{eq:MpartMC2},\\
    P_{3}=&F_{\text{M}}\sin(\theta_2)+(R-V_{\text{EM}})\cos(\theta_2)
    \label{eq:MpartFC3},\\
    V_{3}=&-F_{\text{M}}\cos(\theta_2)+(R-V_{\text{EM}})\sin(\theta_2)
    \label{eq:MpartVC3},\\
    M_{3}=&M_{\text{EM}}-aF_{\text{M}}-bV_{\text{EM}}-F_{\text{M}}x\cos(\theta_2)
    +(R-V_{\text{EM}})x\sin(\theta_2)
    \label{eq:MpartMC3},\\
    P_{4}=&(-R+V_{\text{EM}})\sin(\theta_1)-F_{\text{M}}\cos(\theta_1)
    \label{eq:MpartFC4},\\
    V_{4}=&(-R+V_{\text{EM}})\cos(\theta_1)+F_{\text{M}}\sin(\theta_1)
    \label{eq:MpartVC4},\\
    M_{4}=&M_{\text{EM}}-bV_{\text{EM}}-(R-V_{\text{EM}})x\cos(\theta_1)
    +F_{\text{M}}(x-L_1)\sin(\theta_1)+RL_3\sin(\theta_2)-V_{\text{EM}}L_3\sin(\theta_2).
    \label{eq:MpartMC4}
\end{align}
Thus, the total strain energy of the M part, $SE_{\text{M}}$, is given by,
\begin{align}
SE_{\text{M}}&=\sum_{i=1}^{4}\left[\int\limits_0^{l_i}\left(  \frac{M_{i}^2}{2YI_{\text{M}}} \  
        +\frac{\tau_i^2A_{\text{M}}}{2G}\
        +\frac{\sigma_i^2A_{\text{M}}}{2Y}\right)dx_i\right].
    \label{eq:SEMpart}
\end{align}
Here, $A_{\text{M}}$ is the area, $I_{\text{M}}$ is the second moment of area of the links, $\sigma_i$ and $\tau_i$ are the axial stress and shear stresses in the $i^{th}$ link.
The conditions arising from the mechanism's symmetry, on rotation and on vertical displacement at point E is valid in the case of M part too. Thus, we have, 
\begin{align}
    \frac{dSE_{\text{M}}}{dM_\text{EM}} =& 0,~\text{and}~
    \frac{dSE_{\text{M}}}{dV_\text{EM}} = 0,       \\
    \implies
    M_{\text{EM}}=&\alpha_{M_{\text{EM}}}F_{\text{M}}+\beta_{M_{\text{EM}}}R,
    \label{eq:MEMPart}\\
    V_{\text{EM}}=&\alpha_{V_{\text{EM}}}F_{\text{M}}+\beta_{V_{\text{EM}}}R,
    \label{eq:VEMPart}   
\end{align}
where $\alpha_{M_{\text{EM}}}$, $\beta_{M_{\text{EM}}}$, $\alpha_{V_{\text{EM}}}$, and $\beta_{V_{\text{EM}}}$ are coefficients defined for brevity. These dependent on  the geometrical parameters of the mechanism. Substituting  \cref{eq:MEMPart,eq:VEMPart} in \cref{eq:SEMpart}  gives the total strain energy of M part.

Thus, the displacement $\delta_{\text{M}_X}$ in the X-direction due to the force $F{\text{M}}$ is given by,
\begin{align}
    \delta_{\text{M}_X}=\frac{\partial{SE_{\text{M}}}}{\partial{F_{\text{M}}}}
    =A_{X_{F_{\text{M}}}}F_{\text{M}}+C_{X_R}R
    \label{eq:Dispalcement_X_MPart},
\end{align}
and the displacement in the Y-direction at point N, denoted as $\delta_{\text{M}_Y}$, is,
\begin{align}
     \delta_{\text{M}_Y}=\frac{\partial{SE_{\text{M}}}}{\partial{R}}
    =A_{Y_{F_{\text{M}}}}F_{\text{M}}+C_{Y_R}R
    \label{eq:Dispalcement_Y_MPart}.   
\end{align}
Here, $A_{X_{F_{\text{M}}}}, C_{X_R}, A_{Y_{F_{\text{M}}}}$ and $C_{Y_R}$ denote the coefficients of forces 
\vspace{0.25mm}
$F_{\text{M}}$ and $R$, respectively, which are only dependent on the geometrical parameters of the mechanism.

The displacements of the re-entrant points E and N, contributed by the $\Lambda$ and M parts, are expressed in terms of the reaction force $R$ and the components of the applied force, $F_{\Lambda}$ and $F_{\text{M}}$. These force components are determined in the following section using the principles of compatibility and superposition.

\subsubsection{Compatibility and superposition conditions}
The horizontal displacement of point E, $\delta_{\Lambda_X}$,
from the $\Lambda$ part (Eq. \eqref{eq:Dispalcement_X_LambdaPart}) and $\delta_{\text{M}_X}$
from the M part (Eq. \eqref{eq:Dispalcement_X_MPart}), together with the vertical displacement of point N,   $\delta_{\Lambda_Y}$
from the $\Lambda$ part (Eq. \eqref{eq:Dispalcement_Y_LambdaPart}) and $\delta_{\text{M}_Y}$
from the M part (Eq. \eqref{eq:Dispalcement_Y_MPart}), yield two compatibility equations, given by,
\begin{align}
    \delta_{\Lambda_X}=&\delta_{\text{M}_X}
    \label{eq:CompX}, ~\text{and}\\
    \delta_{\Lambda_Y}=&\delta_{\text{M}_Y}.
    \label{eq:CompY}
\end{align}

The combined mechanism is a parallel combination of the $\Lambda$ and M part, therefore, we  use the superposition principle and express the force $F$ applied on the combined mechanism as the sum of the forces on the $\Lambda$ part, $F_{\Lambda}$ and M part, $F_{\text{M}}$, as,
\begin{equation}
    F=F_{\Lambda}+F_{\text{M}}.
    \label{eq:Force}
\end{equation}
Solving \cref{eq:CompX,eq:CompY,eq:Force}
together for the unknown forces, $F_{\Lambda}$, $F_{\text{M}}$ and R, gives them in terms of the applied force, $F$, as below,
\begin{align}
     F_{\Lambda} =& \gamma_{F_{\Lambda}} F
    \label{eq:ForceF1},\\
    F_{\text{M}} =& \gamma_{F_{\text{M}}} F,~\text{and}
    \label{eq:ForceF2}\\
    R =& \gamma_R F.
    \label{eq:ForceR}    
\end{align}
The coefficients $\gamma_{F_{\Lambda}}$, $\gamma_{F_{\text{M}}}$, and $\gamma_R$ are functions of the mechanism’s geometric parameters.

Substituting 
\cref{eq:ForceF2,eq:ForceR} back in  $\delta_{\text{M}_X}$
given by \cref{eq:Dispalcement_X_MPart}, and $\delta_{\text{M}_Y}$, given by
\cref{eq:Dispalcement_Y_MPart}, of M part will give the horizontal and vertical displacements of E and N points of the combined mechanism as,
\begin{align}
    \Delta_X =& A_{X_{F_{\text{M}}}}\gamma_{F_{\text{M}}}F+C_{X_R}\gamma_RF, ~\text{and}
    \label{eq:XDispCombined}\\
    \Delta_Y =& A_{Y_{F_{\text{M}}}}\gamma_{F_{\text{M}}}F+C_{Y_R}\gamma_RF.
    \label{eq:YDispCombined}    
\end{align}
\subsubsection{Stretch ratios}
The stretch ratio of the mechanism, $\mu$, as defined earlier, is given by the ratio of $\Delta_X$  to $2 \Delta_Y$. That is,
\begin{equation}
    \mu=\frac{\Delta_X}{2\Delta_Y} \label{eq:SR}.
\end{equation}
The factor of 2 in the denominator accounts for the fact that the force $F$ applied in \cref{fig:Geometrical parameters} is twice the force $f$ used in \cref{fig:Mechanism with links as beams}, due to the symmetry-based simplification. 
Since both 
$\Delta_X$ and $\Delta_Y$ vary linearly with the applied force $F$, their ratio $\mu$ is independent of $F$. Therefore, $\mu$ depends solely on the geometric parameters of the mechanism (see Fig.~\ref{fig:Geometrical parameters}).

\Cref{eq:SR} defines the stretch ratio $\mu$ of the combined mechanism. The mechanism exhibits a negative stretch ratio, denoted by $\mu_N$, when the inner $\Lambda$ part is present, i.e., when $t_{\Lambda}>0$ and $h_{\Lambda}>0$. Conversely, when the inner $\Lambda$ part is absent (see \cref{fig:M and V part}), effectively modeled with zero stiffness $k=0$ corresponding to $t_{\Lambda} \approx 0$ and $h_{\Lambda} \approx 0$, the mechanism exhibits a positive stretch ratio, denoted by $\mu_P$. Thus, from \cref{eq:SR}, we define,
\begin{align}
        \mu_N=\left(\frac{\Delta_X}{2\Delta_Y}\right)_{t_{\Lambda}>0,~h_{\Lambda}>0},
    \label{eq:SRN}\\
        \mu_P=\left(\frac{\Delta_X}{2\Delta_Y}\right)_{t_{\Lambda}\approx0,~h_{\Lambda}\approx0}
    \label{eq:SRP}.
\end{align}

Here, it turns that $\mu_P$ is primarily governed by the angular parameter $\theta_1$ and weakly dependent on  on other geometric parameters is relatively minor.
 In contrast, $\mu_N$ is strongly dependent on the geometric dimensions $t_{\Lambda}$, $h_{\Lambda}$, $t_{\text{M}}$, $h_{\text{M}}$, $a$, and $b$ (see \cref{fig:Geometrical parameters}) of the combined mechanism. In our design approach, we first fix $\mu_P$ by selecting $\theta_1$, and then tune the remaining parameters to achieve the desired $\mu_N$.

\subsection{Modeling for tuning stiffness}
\begin{figure*}[h]
    \centering
    \includegraphics[width=1\linewidth]{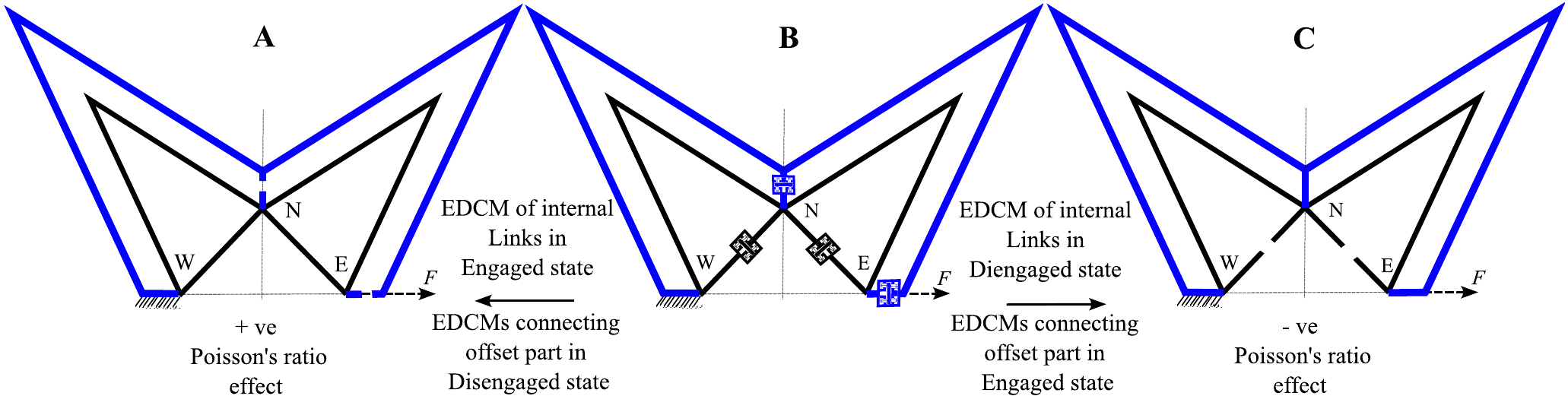}
    \caption{ (A) Effective mechanism when it is in the negative stretch state, here the offset mechanism(in blue color) is not connected to the main frame (in black color). (B) Mechanism with EDCM to facilitate the engagement and disengagement of the offset mechanism. (C) Effective mechanism in the positive stretch state, where the offset mechanism is connected to the initial mechanism.}
    \label{fig:IllustrationStiffness}
\end{figure*}

The effective stiffness, $K$, of the entire mechanism between points E and W (see \cref{fig:Geometrical parameters}) is defined as the ratio of the applied force $F$ to the resulting displacement $\Delta_X$, and is given by:
\begin{equation}
    K=\frac{F}{\Delta_X}.
    \label{eq:stifness}
\end{equation}
As mentioned before, this stiffness changes when the mechanism switches between positive and negative stretch ratios configurations. And equalization of these stiffness is achieved via the engagement of the offset mechanism via EDCMs, as illustrated in \cref{fig:IllustrationStiffness}.

The stiffness in the negative stretch ratio state (positive Poisson's effect), $K_n$, can be obtained as, 
\begin{equation}
    K_n=\frac{F}{\Delta_X}.
    \label{eq:StiffnesskN}
\end{equation}

To get the stiffness in the positive stretch state, $K_p$,  we need to find the displacement of point E along $F$ for the mechanism with offset part engaged (see \cref{fig:IllustrationStiffness}C).
 By varying the dimension of the offset mechanism, we can vary the stiffness of the positive stretch ratio case (negative Poisson's effect).
 
\begin{figure*}[h]
    \centering
    \includegraphics[width=\linewidth]{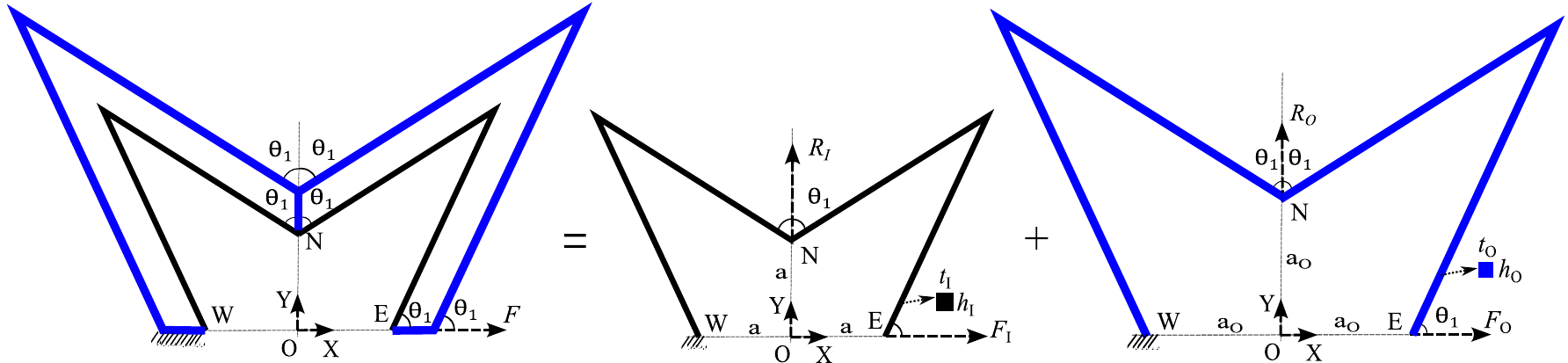}
    \caption{Mechanism in the positive stretch state split as a parallel combination of MI(Black) and MO(Blue) parts}
    \label{fig:SplittedM}
\end{figure*}

 Next, we derive the relation between the stiffness and geometrical parameters of the offset part by splitting the mechanism into two parts:  an offset M-shaped frame (referred to as the MO part) and an inner M-shaped frame (referred to as the MI part) as shown in \cref{fig:SplittedM}.
\subsubsection{Compatibility and superposition conditions}
\Cref{eq:Dispalcement_X_MPart} gives the horizontal displacement $\delta_{\text{M}_X}$ of the M part under an applied horizontal force. This relation is applicable individually to both the MO and MI parts. By substituting $F_{\text{I}}$ and $R_{\text{I}}$ (see \cref{fig:SplittedM}) in place of $F_{\text{M}}$ and $R$, the displacement of the MI part becomes,
\begin{align}
\delta_{\text{MI}_X} = A_{X_{F_{\text{I}}}} F_{\text{I}} + C_{X_{R_{\text{I}}}} R_{\text{I}}. \label{eq:DispXMI}
\end{align}
Similarly, by replacing the forces and geometric parameters of $\delta_{\text{M}_X}$ with those of the MO part—i.e., $a_{\text{O}}$, $h_{\text{O}}$, $t_{\text{O}}$, $F_{\text{O}}$, and $R_{\text{O}}$—we obtain the horizontal displacement of the MO part as,
\begin{align}
\delta_{\text{MO}_X}=A_{X_{F_{\text{O}}}}F_{\text{O}}+C_{X_{R_\text{O}}}R_\text{O}.\label{eq:DispXMO}
\end{align}
The corresponding vertical displacements of the MI and MO parts are,
\begin{align}
\delta_{\text{MI}_Y}=A_{Y_{F_{\text{I}}}}F_{\text{I}}+C_{Y_{R_{\text{I}}}}R_{\text{I}}~\text{and}\label{eq:DispYMI}\\    \delta_{\text{MO}_Y}=A_{Y_{F_{\text{O}}}}F_{\text{O}}+C_{Y_{R_\text{O}}}R_\text{O}.\label{eq:DispYMO}   
\end{align}
When the MI and MO parts are combined to get the desired mechanism,  the following compatibility and equilibrium conditions must be satisfied:
\begin{align}
\delta_{\text{MO}_X}=\delta_{\text{MI}_X}, \label{eq:compatX}\\
\delta_{\text{MO}_Y}=\delta_{\text{MI}_Y}, \label{eq:compatY}\\
F=F_\text{I}+F_\text{O}, ~ \text{and} \label{eq:forceBalance}\\
R_{\text{I}}+R_\text{O}=0. \label{eq:reactionBalance}
\end{align}Solving the system of  \cref{eq:compatX,eq:compatY,eq:forceBalance,eq:reactionBalance} yields the internal forces $F_{\text{I}}$, $R_{\text{I}}$, $F_{\text{O}}$, and $R_{\text{O}}$ in terms of the externally applied force $F$:
\begin{align}
     F_{\text{O}} =& \gamma_{F_{\text{O}}} F
    \label{eq:ForceFO},\\
    F_{\text{I}} =& \gamma_{F_{\text{I}}} F,
    \label{eq:ForceFI}\\
    R_{\text{I}} =& \gamma_{R_{\text{I}}} F, ~\text{and}\label{eq:RI}\\
    R_\text{O}=&-R_{\text{I}}.
    \label{eq:ForcesDM}    
\end{align}
Substituting the expressions for $F_{\text{I}}$ and $R_{\text{I}}$ into  \cref{eq:DispXMI}  gives the horizontal displacement
of the whole mechanism, \begin{equation}
   \Delta_{\text{M}_X} = A_{X_{F_{\text{\text{I}}}}}\gamma_{F_{\text{\text{I}}}}F+C_{X_{R_I}}\gamma_{R_I}F.
\end{equation} The stiffness in the positive stretch configuration, $K_p$, 
can then be obtained as:\begin{align}
    K_p=&\frac{F}{\Delta_{\text{M}_X}}.\label{eq:StiffnessKp}
\end{align} 

When the dimensions of the MI part are fixed, the stiffness in the positive stretch ratio state, $K_p$, can be tuned by varying the geometric parameters of the MO part namely, $a_{\text{O}}$, $h_{\text{O}}$, and $t_{\text{O}}$ (see \cref{fig:SplittedM}).

\section{Design and illustrative examples}\label{sec:Design}
While the stretch ratio and stiffness expressions derived in the previous sections are valid for asymmetric substrates as well, we hereafter assume geometric symmetry about the cell adhesion points by setting $a = b$ and $\theta_1 = \theta_2$ (see \cref{fig:Geometrical parameters}).
\subsection{Designing stretch ratio}\label{sec:Design for SR}

As mentioned earlier, $\mu_P$ can be effectively tuned by varying the angle $\theta_1$, since for practical values of $\mu_P$ (up to approximately 4), its dependence on other geometric parameters is relatively minor. This allows us to begin with an initial guess for the remaining parameters, adjust $\theta_1$ to achieve the desired $\mu_P$, and later revisit $\mu_P$ once the full geometry is finalized based on $\mu_N$. 

While varying $\theta_1$ to tune $\mu_P$, a lower bound must be imposed on $\theta_1$ to ensure that the mechanism intersects within the first quadrant, which requires $\theta_1 + \theta_2 > 90^\circ$. Under the symmetry condition $\theta_1 = \theta_2$, this leads to the constraint $\theta_1 > 45^\circ$. Also, the geometry approaches a square configuration and loses this characteristic when $\theta_1$ nears $90^\circ$.  So, an upper limit of $70^\circ$ is imposed to maintain the re-entrant nature of the geometry.

\begin{figure}[!htbp]\centering
    \includegraphics[width=1\linewidth]{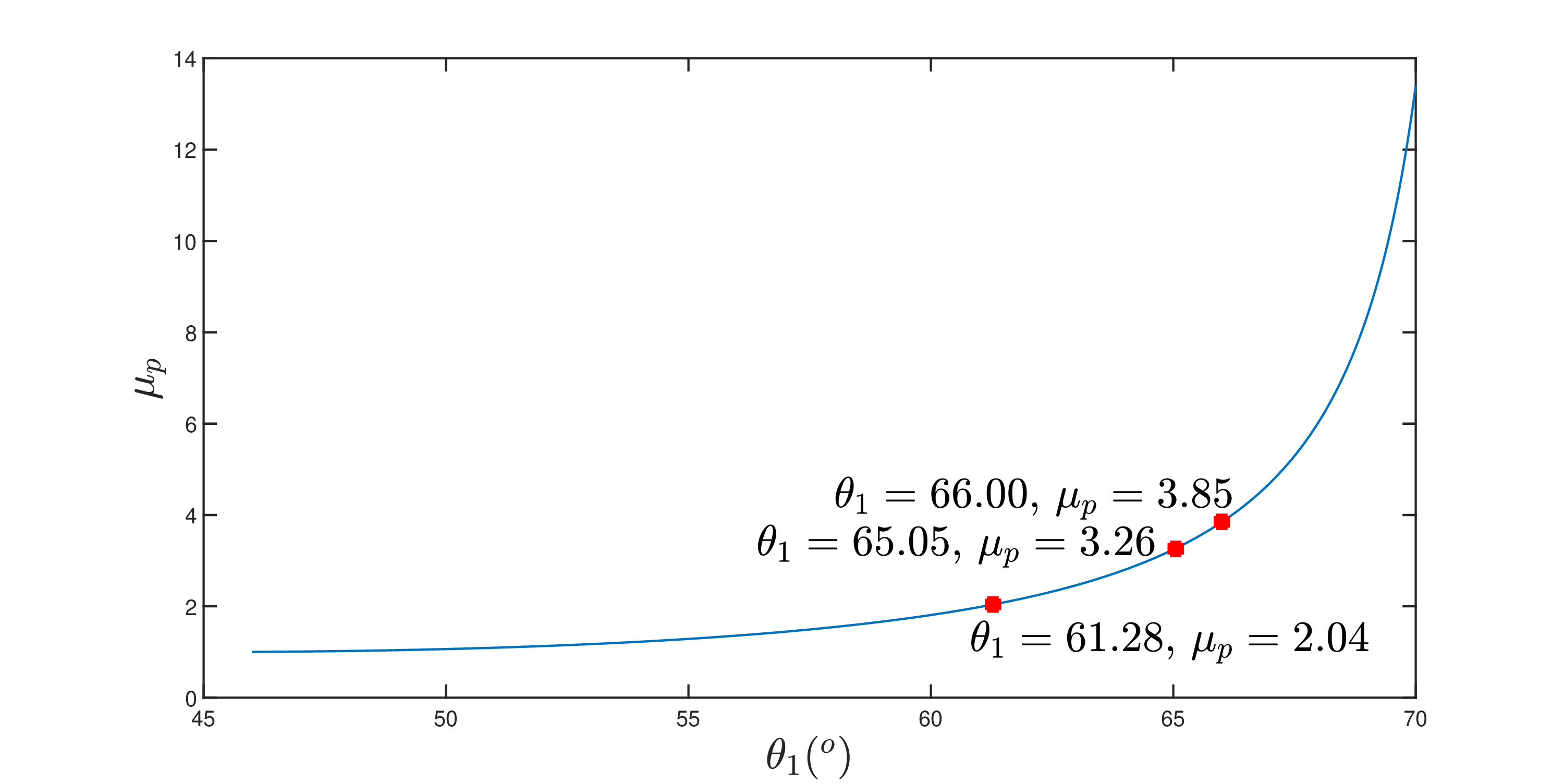}
    \caption{Design curve illustrating the dependence of $\mu_P$ on $\theta_1$ ($\theta_2 = \theta_1$). 
    }
    \label{fig:SRP Curve}
\end{figure}

By using these bounds, and by fixing the other dimensions to some arbitrary values (as their influence is minimal), a design curve for $\mu_P$ is generated
from \cref{eq:SRP}, as shown in \cref{fig:SRP Curve}. As illustrated, varying $\theta_1$ from $45^\circ$ to $70^\circ$ allows $\mu_P$  to be tuned within a range of approximately 1 to 10.

Once a value of $\theta_1$ is selected for a desired $\mu_P$, the parameters $t_{\Lambda}$, $h_{\Lambda}$, $t_{\text{M}}$, $h_{\text{M}}$, and $a$  can be used to design   $\mu_n$. To simplify the design process, we fix the out-of-plane thicknesses of both the $\Lambda$  and M parts based on the fabrication constraints.  
The in-plane thickness of the M part is fixed according to the value of the slenderness ratio, $s_\text{M}$, considered. 

\begin{figure}[!htbp]
    \centering
    \includegraphics[width=\linewidth]{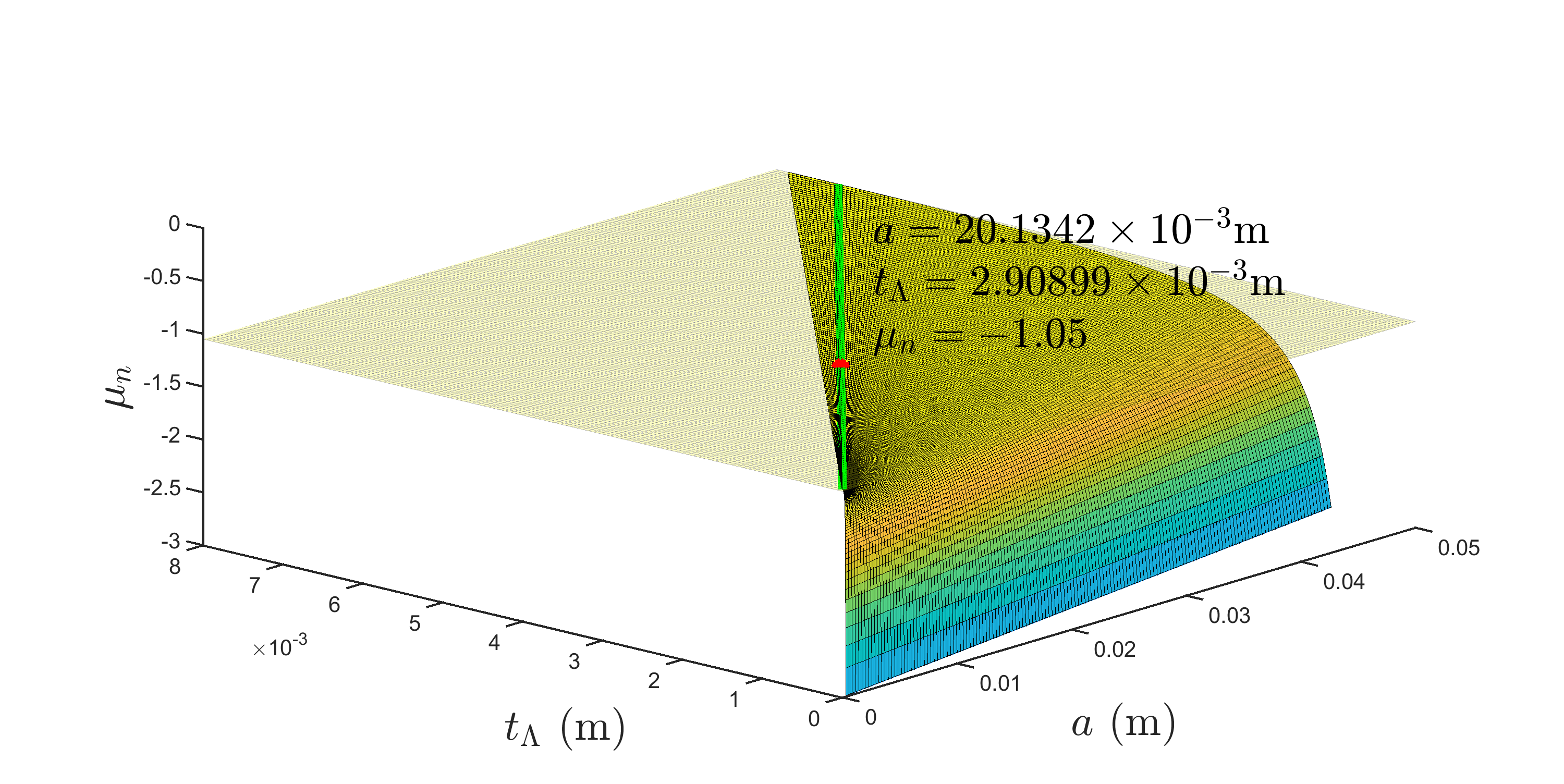}
    \caption{Design surface illustrating the achievable range of $\mu_n$ as a function of $a$ and $t_{\Lambda}$ for Example 1.}
    \label{fig:Example1-Contour}
\end{figure}

With these assumptions, the stretch ratio $\mu_n$ becomes a function only of the parameters $a$ and $t_{\Lambda}$. By varying these two parameters, a design surface for $\mu_n$ can be generated, as shown in  \cref{fig:Example1-Contour}.. The maximum permissible value of $a$ is often limited by  fabrication constraints. The maximum value of $t_{\Lambda}$
is limited 
by the minimum value of the slenderness ratio, $s_\Lambda$, considered for that part.

The design curve for $\mu_P$ and the design surface for $\mu_N$ together enable the design of the mechanism with desired stretch ratios. This design process is illustrated through an example next.
\subsubsection{Example 1: $\mu_N=-1.05$, $\mu_P=3.26$}
 In Example 1, we consider the design of a mechanism with a $\mu_p=3.26$ 
 and $\mu_n=-1.05$. To begin, we select the angle corresponding to the desired $\mu_P$ from the design curve (see \cref{fig:SRP Curve}). We plot this curve by guessing $a=10\times10^{-3}$ m, $t_{M}=1\times10^{-3}$ m, and $h=2\times10^{-3}$ m. For $\mu_P = 3.26$, the required angle is $\theta_1 = 65.05^\circ$. With this selected value of $\theta_1$ and  out-of-
plane thicknesses $h_{\Lambda}=h_{\text{M}}=2\times10^{-3}$ m, the design surface for $\mu_n$ is generated by varying the parameter $a$ from 0 to 40 mm, 
 and $t_{\Lambda}$ from 0 to $t_{\Lambda}=\frac{L_1}{s_{\Lambda}}$.
 The slenderness ratios assumed for the $\Lambda$ and M parts are $s_{\Lambda} = 9$ and $s_{\text{M}} = 10$, respectively.   To identify configurations yielding the desired $\mu_n = -1.05$, we intersect the design surface with a horizontal plane at $\mu_n = -1.05$. This intersection produces a design curve, shown as the green curve in \cref{fig:Example1-Contour}.

Each point on this curve represents a feasible design solution. For demonstration, we select one such point (indicated by the marker),  where $a = 20.134\times10^{-3}$ m and $t_{\Lambda}=2.908\times10^{-3}$ m. To validate the accuracy of $\mu_P$, we substituted the arbitrary parameters used in its initial design curve with the newly derived values. The absence of any notable change confirmed its reliability, thus completing the design process.
The geometrical parameters corresponding to this model are tabulated under Example 1 in \cref{tab: Geometrical parameters}. 
\begin{table}[!htbp]
\centering
\setlength{\tabcolsep}{0.6\tabcolsep}
{\begin{tabular}{|c|c|c|c|c|c|c|c|}
\hline
Ex. & \begin{tabular}{@{}c@{}}$a$ \\(mm) \end{tabular} & \begin{tabular}{@{}c@{}}$a_{\text{O}}$ \\(mm) \end{tabular} & \begin{tabular}{@{}c@{}}$\theta_1$ \\($^o$) \end{tabular} & \begin{tabular}{@{}c@{}}$h$ \\(mm) \end{tabular} & \begin{tabular}{@{}c@{}}$t_{\text{M}}$ \\(mm) \end{tabular}    & \begin{tabular}{@{}c@{}}$t_{\Lambda}$ \\(mm) \end{tabular}    & \begin{tabular}{@{}c@{}}$t_{\text{O}}$ \\(mm) \end{tabular}    \\ \hline
1       & 20.134 & -      & 65.05  & 2     & 4.152 & 2.908 & -     \\ \hline
2       & 15     & 32.060 & 61.28  & 2     & 2.615 & 1.192 & 9.095 \\ \hline
3       & 115    & -      & 66     & 5     & 5     & 5     & -     \\ \hline
\end{tabular}}
\caption{Geometrical parameters of Examples 1, 2, and 3. }
\label{tab: Geometrical parameters}
\end{table}

\subsection{Designing  stretch ratio considering stiffness}\label{sec:designforstiffness}

\begin{figure*}[h]
    \centering
    \includegraphics[width=0.7\linewidth]{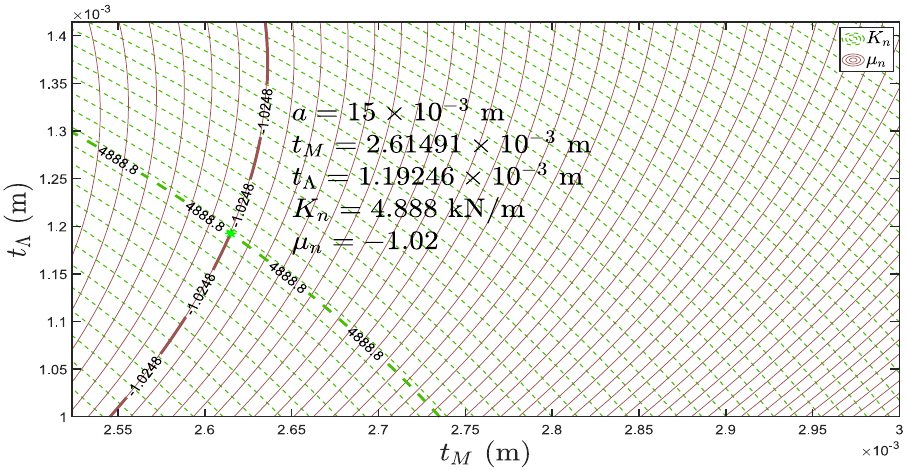}
    \caption{Design contour plot used to select suitable values of $t_{\Lambda}$ and $t_\text{M}$ that achieve the desired stretch ratio $\mu_n$ and stiffness $K_n$.
    }
    \label{fig:StretchStiffnessContour}
\end{figure*}

We design $\mu_p$ as before. By assuming  out-of-plane thicknesses and fixing $a$ based on fabrication constraints, a contour plot is generated showing $\mu_n$ and $K_n$ as functions of the remaining in-plane thicknesses, $t_{\Lambda}$ and $t_\text{M}$, using \cref{eq:SRN,eq:StiffnesskN}. The intersection of the contours corresponding to the desired $\mu_n$ and $K_n$ gives the required values of $t_{\Lambda}$ and $t_\text{M}$ (see \cref{fig:StretchStiffnessContour}).  

\begin{figure}[!htbp]
    \centering
    \includegraphics[width=\linewidth]{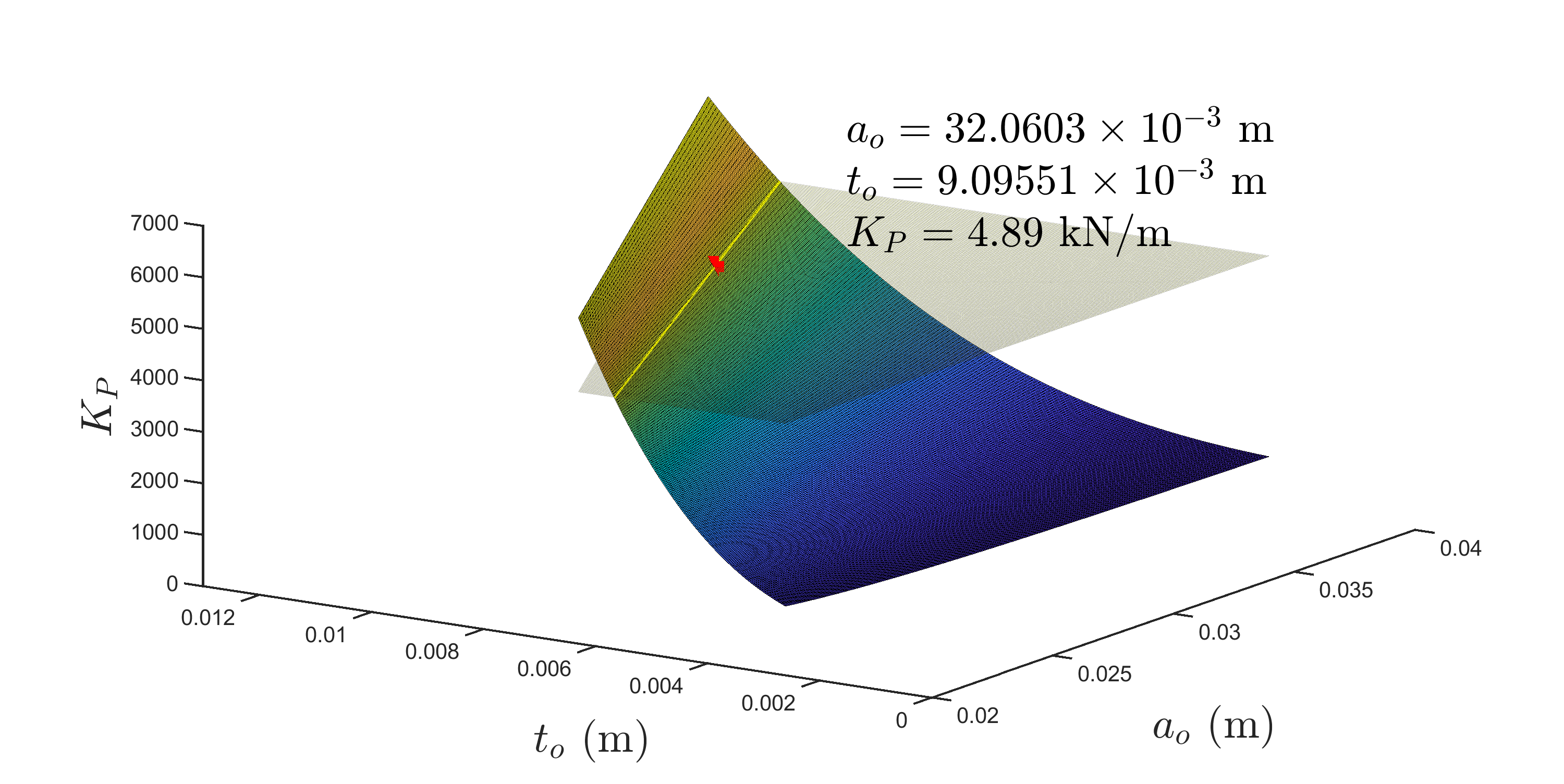}
    \caption{Design surface for the positive stretch stiffness, $K_p$, generated for Example 2 by varying the MO part dimensions $a_{\text{O}}$ and $t_{\text{O}}$.}
    \label{fig:StiffnessKp}
\end{figure}

To achieve the desired stiffness in the positive stretch configuration, the MO segment is engaged. The required dimensions of the MO part are determined from a design surface of $K_p$ generated by varying $a_{\text{O}}$ and $t_{\text{O}}$ in \cref{eq:StiffnessKp} (see \cref{fig:StiffnessKp}).
 Intersecting this surface with a horizontal plane at the target stiffness yields a design curve (highlighted in yellow). A suitable point on this curve is selected based on fabrication feasibility, thus fixing $a_{\text{O}}$ and $t_{\text{O}}$ and completing the design. This procedure is illustrated in the following example.

%%%%%%%%%%%%%%%%%%%%%%%%%%%%%%%%%%%%%%%%%%%%%%%%%%%%%%%%%%%%%%%%%%%%%
\subsubsection{Example 2: $\mu_n=-1.02$, $\mu_p=2.04$ and $K=4.89$ kN/m}
In this example, we intend to design  $\mu_p=2.04$. From the design curve in \cref{fig:SRP Curve}, this corresponds to $\theta_1 = 61.28^\circ$.  For ease of fabrication, we set   $a = 15\times10^{-3}$ m and fix all out-of-plane thickness at $h=2\times10^{-3}$ m. 
To realize $\mu_n = -1.02$ with a stiffness of $4.89~\text{kN/m}$, we identify  the corresponding contour lines in  \cref{fig:StretchStiffnessContour} and obtain the required values of $t_{\Lambda}$ and $t_{\text{M}}$.

With these values, the initial stiffness in the positive stretch ratio configuration is $0.9 \text{~kN/m}$. To raise this to the desired $4.89$ kN/m, we design the MO segment using the design surface in \cref{fig:StiffnessKp}. Choosing the marked point on the curve yields $a_{\text{O}} = 32.06\times10^{-3}$ m and $t_{\text{O}} = 9.095\times10^{-3}$ m. The complete set of geometrical parameters is summarized in \cref{tab: Geometrical parameters}. Validation of this design and Example 1 follows in the next section.

\section{Validation}\label{sec:Validation}
\begin{figure}[!htbp]
    \centering
    \includegraphics[width=0.5\linewidth]{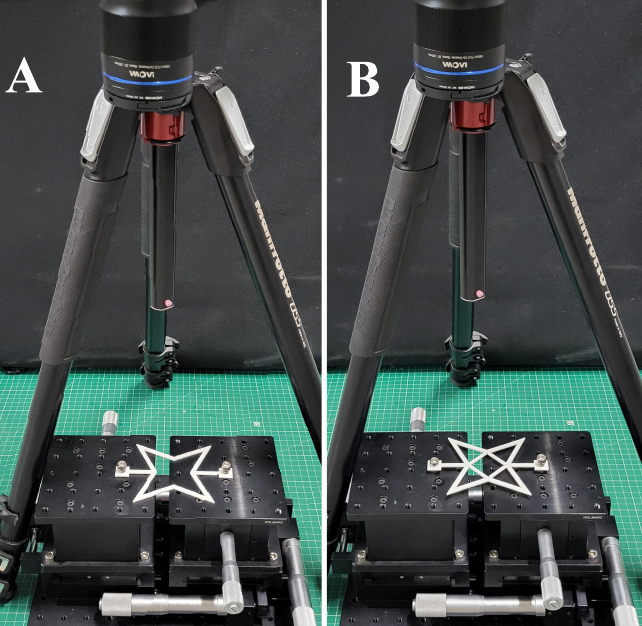}
     \caption{Experimental setup used to validate Example 1. (A) Configuration for testing the positive stretch ratio.
     (B) Configuration for testing the negative stretch ratio. 
     }
    \label{fig:SREXPSetup}
\end{figure}

The positive and negative stretch ratio states are validated using separate physical prototypes, shown in \cref{fig:SREXPSetup}. The positive stretch state (EDCM disengaged) is represented by a mechanism without the interconnecting link (\cref{fig:SREXPSetup}A), while the negative stretch state (EDCM engaged) uses a mechanism with the link in place (\cref{fig:SREXPSetup}B).These designs are also validated through FEA. Validation of the full mechanism with integrated EDCMs is presented later in \cref{sec:EDCM}.

\subsection{Validation of Example 1}

\subsubsection{Experiment}\label{subsec:SRExp}
The mechanism was fabricated using an FDM 3D printer
with Hyper PLA filaments.
The material has a density of 1240 kg/m\textsuperscript{3}, a Young’s modulus of 3.576 GPa, and a Poisson’s ratio of 0.22. Extensions were added at points E and W (see \cref{fig:Geometrical parameters}) to facilitate connecting to linear stages for experimental testing. During the experiment, the linear stage attached to the left end of the mechanism (point W) was fixed, while the stage on the right (point E) was actuated to apply displacement (see \cref{fig:SREXPSetup}).

Top-view images of the mechanism before and after deformation were captured and overlaid. Displacements at points E and N were extracted using ImageJ \cite{imagej} (see Figs.~\ref{fig:Ex1FEAExp}C and D). The stretch ratio was computed using \cref{eq:SR}, and the  measured displacements and corresponding stretch ratios are summarized in \cref{tab:Exp FEA values SR}.

\begin{table*}[h]
\centering
\renewcommand{\arraystretch}{1.5}
\resizebox{\textwidth}{!}{\begin{tabular}{|c|l|l|l|l|l|l|l|l|l|l|l|}
\hline
\multicolumn{1}{|l|}{Ex. No.} & $\mu$ & Force (N) & \begin{tabular}{@{}c@{}}$\Delta_X$ ($\times10^{-3}$m) \\ FEM\end{tabular} & \begin{tabular}{@{}c@{}}$\Delta_Y$ ($\times10^{-3}$m) \\ FEM\end{tabular}& \begin{tabular}{@{}c@{}}$\Delta_X$ ($\times10^{-3}$m) \\ Analytical\end{tabular}& \begin{tabular}{@{}c@{}}$\Delta_Y$ ($\times10^{-3}$m) \\ Analytical\end{tabular} & \begin{tabular}{@{}c@{}}$\Delta_X$ ($\times10^{-3}$m) \\ Experiment\end{tabular} & \begin{tabular}{@{}c@{}}$\Delta_Y$ ($\times10^{-3}$m) \\ Experiment\end{tabular} & \begin{tabular}{@{}c@{}} $\mu$\\ FEM\end{tabular} & \begin{tabular}{@{}c@{}}$\mu$ \\ Analytical\end{tabular} & \begin{tabular}{@{}c@{}}$\mu$ \\ Experiment\end{tabular} \\ \hline
\multirow{2}{*}{1} & $\mu_P$ & $3.8055 $& $1.4845$ & $0.2271$ &  $1.4821$ &  $0.2274$ &  $1.4815$ &  $0.2222$ & $3.2684$ & $3.2588$ & $3.3337$ \\ \cline{2-12} 
 & $\mu_N$ & $15.0597$ & $1.0370$ & $-0.4980$ & $1.0374$ &$-0.4982$& $1.0370$ & $-0.5000$& $-1.0412$ & $-1.0411$ & $-1.0370 $\\ \hline
\end{tabular}}
\caption{Comparison of stretch ratio results from finite element analysis (FEM), analytical model, and experiments for Example 1.}
\label{tab:Exp FEA values SR}
\end{table*}
 
\subsubsection{FEA} \label{subsec:FEASR}
\begin{figure*}[!htbp]
    \centering
    \includegraphics[width=0.6\linewidth]{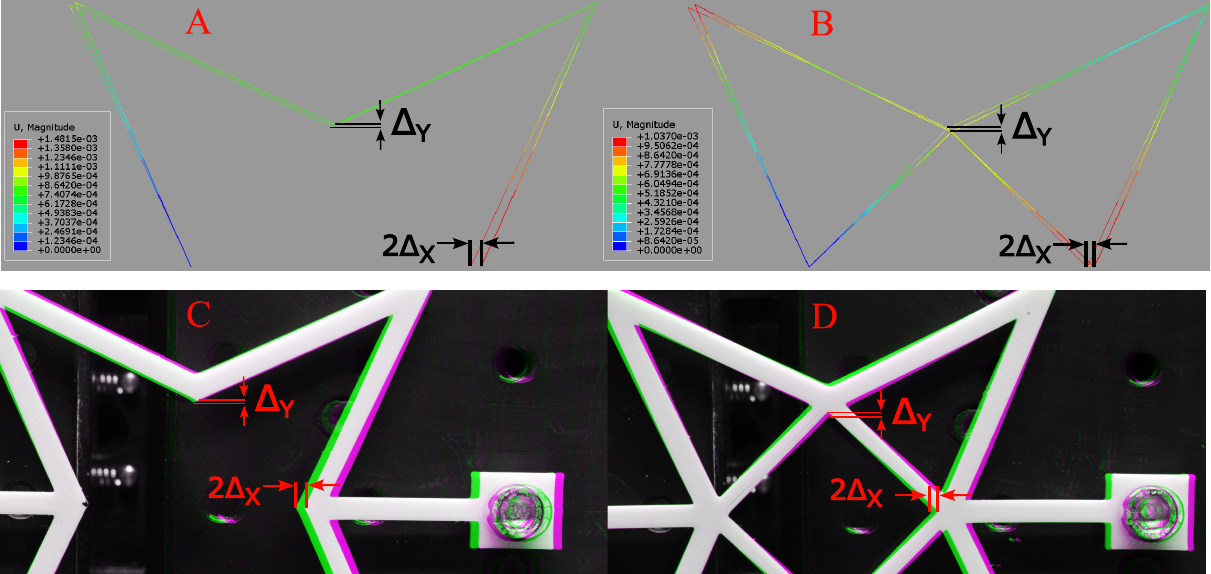}
    \caption{Deformed mechanisms superimposed over initial configurations from FEA and experiments for Example 1. (A) Positive stretch state from FEM; (B) Negative stretch state from FEM; (C) Positive stretch state from experiment (green: initial, pink: deformed, white: overlap); (D) Negative stretch state from experiment.}
    \label{fig:Ex1FEAExp}
\end{figure*}

A quasi-static analysis of the mechanism was done in Abaqus 2018 \cite{abaqus2018software} using a linear wire model (NLGEOM turned off) with beam elements and PLA material properties.  
The mesh size was set to $1\times10^{-5}$ (see Figs. \ref{fig:Ex1FEAExp} A and B). A prescribed horizontal displacement, matching the experimental setup, was applied at point E, and the resulting vertical displacement and reaction force were extracted.

 The corresponding stretch ratios, $\mu$, for both positive and negative stretch states were then calculated using \cref{eq:SR}. Analytical values of $\mu$ were also computed by substituting the FEA-extracted force into the analytical expressions for displacement and stretch ratio(see \cref{eq:XDispCombined,eq:YDispCombined,eq:SRP,eq:SRN}). These results are summarized in \cref{tab:Exp FEA values SR}, and a comparison of analytical, FEA, and experimental values is presented in \cref{tab: ComparisonFEAAndExp1}.

The small discrepancies observed, primarily between analytical and experimental values, are likely due to manual setup errors and image scale calibration in ImageJ. Nonetheless, all errors remain within 5\%, confirming strong agreement across all methods.
\begin{table*}[h]
\centering
\renewcommand{\arraystretch}{1.5}
\resizebox{\textwidth}{!}{\begin{tabular}{|c|c|c|c|c|c|c|c|}
\hline
Ex. No &  $\mu$ &  \begin{tabular}{@{}c@{}}\%Error \\ $\Delta_X$ An vs FEM\end{tabular}  & \begin{tabular}{@{}c@{}}\%Error \\ $\Delta_Y$ An vs FEM\end{tabular}   & \begin{tabular}{@{}c@{}}\%Error \\ $\mu$  An vs FEM\end{tabular}  &\begin{tabular}{@{}c@{}}\%Error \\  $\Delta_X$ An vs Exp\end{tabular} &\begin{tabular}{@{}c@{}}\%Error \\$\Delta_Y$ An vs Exp \end{tabular}  & \begin{tabular}{@{}c@{}}\%Error \\  $\mu$ An vs Exp\end{tabular} \\ \hline
\multirow{2}{*}{1} &  $\mu_P$ & 0.1619 & 0.1319 & 0.2942 & 0.0405 & 2.2867 & 2.2988\\ \cline{2-8} 
 &  $\mu_N$ & 0.0386 & 0.0401 & 0.0016 & 0.0386 & 0.3613
 & 0.3984\\ \hline
\end{tabular}}
\caption{Comparison of analytical (An), FEA, Experiment(Exp) results for Example 1}
\label{tab: ComparisonFEAAndExp1}
\end{table*}
%%%%%%%%%%%%%%%%%%%%%%%%%%%%%%%%%%%%%%%%%%%%%%%%%%%%%%%%%
\subsection{Validation of Example 2}
\begin{figure}[!htbp]
    \centering
    \includegraphics[width=0.5\linewidth]{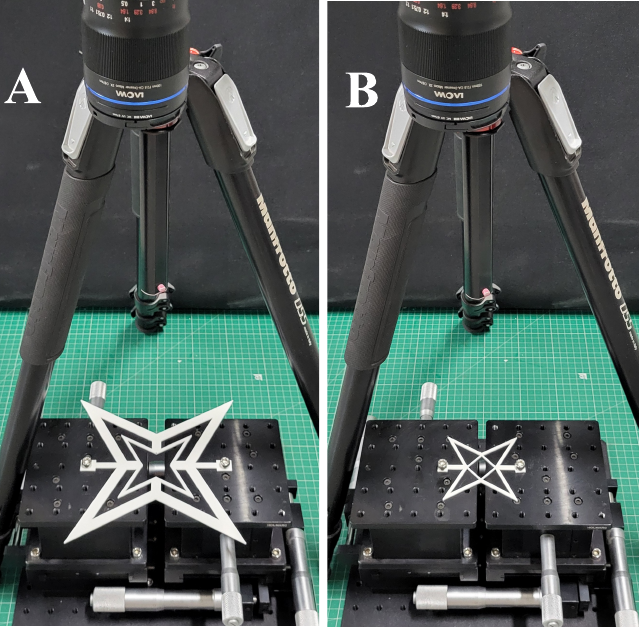}
    \caption{Experimental setup used to validate Example 2. (A) Configuration for testing the positive stretch ratio, with rigid links connecting the MI and MO parts. (B) Configuration for testing the negative stretch ratio.} 
    \label{fig:StiffnessEXPSetup}
\end{figure}

Example 2 is additionally tuned for  a specific stiffness. In the positive stretch ratio state, the inner MI part and the offset MO part are connected via rigid links to represent the engaged state of the EDCM in those regions (see \cref{fig:StiffnessEXPSetup} A). In the negative stretch state, the MO part is omitted, as it is inactive during this state as shown  in (see \cref{fig:StiffnessEXPSetup} B).

%%%%%%%%%%%%%%%%%%%%%%%%%%%%%%%%%%%%%%%%%%%%%%%%%%%%%%%%%%%%%%%%%%%%%%%%%%%%%%%%%%%%%%%
\subsubsection{Experiment}
\begin{figure*}[!htbp]
    \centering
    \includegraphics[width=0.6\linewidth]{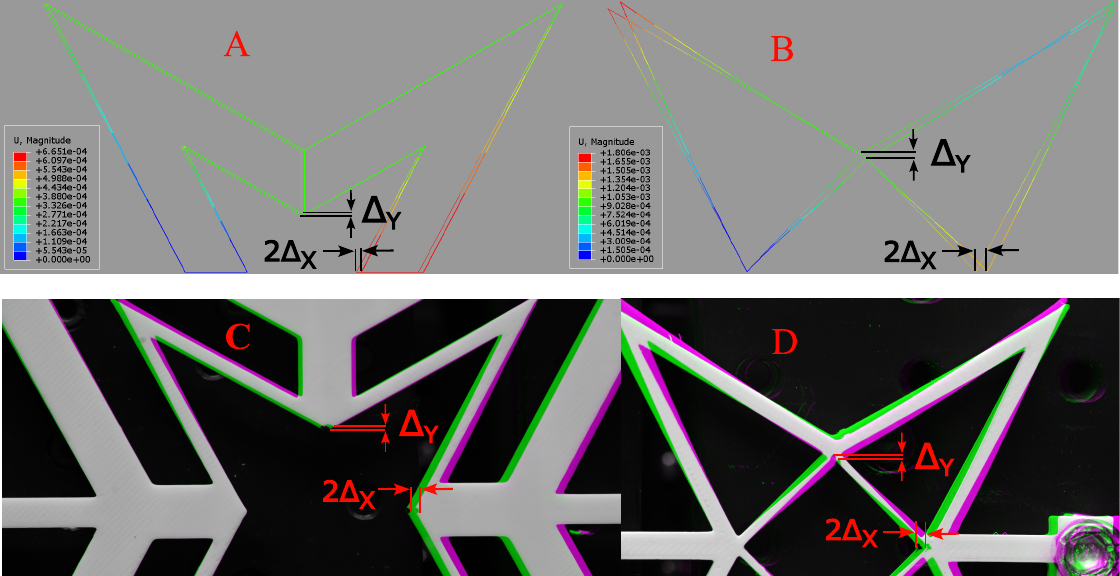}
    \caption{
    Deformed mechanisms superimposed over initial configurations from FEA and experiments for Example 2. (A) Positive stretch state from FEM; (B) Negative stretch state from FEM; (C) Positive stretch state from experiment (green: initial, pink: deformed, white: overlap); (D) Negative stretch state from experiment.}
    \label{fig:Ex2FEAExp}
\end{figure*}

The displacement values extracted using ImageJ (see \cref{fig:Ex2FEAExp} C, D)  are tabulated in \cref{tab:Example2FEAExpANValues}.
\begin{table*}[h]
\centering
\renewcommand{\arraystretch}{1.5}
\resizebox{\textwidth}{!}{\begin{tabular}{|c|l|l|l|l|l|l|l|l|}
\hline
\text{Ex. No} & $\mu$ & \text{Force} (N) & \begin{tabular}{@{}c@{}}$\Delta_X$ ($\times10^{-3}$m) \\ FEM\end{tabular} & \begin{tabular}{@{}c@{}}$\Delta_Y$ ($\times10^{-3}$m) \\ FEM\end{tabular} & \begin{tabular}{@{}c@{}}$\Delta_X$ ($\times10^{-3}$m) \\ Analytical\end{tabular} & \begin{tabular}{@{}c@{}}$\Delta_Y$ ($\times10^{-3}$m) \\ Analytical\end{tabular} & \begin{tabular}{@{}c@{}}$\Delta_X$ ($\times10^{-3}$m) \\ Experiment\end{tabular} &\begin{tabular}{@{}c@{}}$\Delta_Y$ ($\times10^{-3}$m) \\ Experiment\end{tabular} \\
\hline
\multirow{2}{*}{2} & $\mu_p$ & 3.2543 & $0.6651$ & $0.1624$ & $0.6649$ & $0.1625$ & $0.6651$ & $0.1581$ \\ \cline{2-9}
  & $\mu_n$ & $7.1557$ & $1.4636$ & $-0.7140$ & $1.4637$ & $-0.7141$ & $1.4636$ & $-0.7395$ \\
\hline
\end{tabular}}
\caption{Results from FEM, Analytical, and Experimental for Example 2}
\label{tab:Example2FEAExpANValues}
\end{table*}
The corresponding stretch ratios, $\mu$, and stiffness values, $K$, computed using \cref{eq:SR,eq:stifness}, are presented in \cref{tab:Example2FEAExpANSRAndStiffness}. The results show that the stiffness values for the positive and negative stretch states are now nearly equal, demonstrating the effectiveness of the design in balancing stiffness across both deformation modes. 
\begin{table*}[h]
\centering
{\begin{tabular}
{|c|l|l|l|l|l|l|l|}
\hline
\begin{tabular}{@{}c@{}}Ex. \\No. \end{tabular} & $\mu$ &  \begin{tabular}{@{}c@{}}$\mu$ \\ FEM\end{tabular} & \begin{tabular}{@{}c@{}}$\mu$ \\ Analytical\end{tabular} & \begin{tabular}{@{}c@{}}$\mu$ \\ Experiment\end{tabular} & \begin{tabular}{@{}c@{}}$K$(kN/m) \\ FEA\end{tabular} & \begin{tabular}{@{}c@{}}$K$(kN/m) \\ Analytical\end{tabular} &\begin{tabular}{@{}c@{}}$K$(kN/m) \\ Experiment\end{tabular} \\
\hline
\multirow{2}{*}{2} & $\mu_p$ & $2.0169$ & $2.0458$ & $2.1034$ & $4.9677$ & $4.8945$ & $4.8930$ \\ \cline{2-8}
  & $\mu_n$ & $-1.02493$ & $-1.02486$ & $-0.9896$ & $4.8891$ & $4.8888$ & $4.8891$ \\
\hline
\end{tabular}}
\caption{Calculated values of $\mu$ and $K$ of FEM, Analytical, and Experiment for Example 2}
\label{tab:Example2FEAExpANSRAndStiffness}
\end{table*}
%%%%%%%%%%%%%%%%%%%%%%%%%%%%%%%%%%%%%%%%%%%%%%%%%%%%%%%%%%%%%%%%%%%%%%%%%%%%%%%%%%%%%%%%%%
\subsubsection{FEA}
The values extracted from FEM analysis (see \cref{fig:Ex2FEAExp}) is presented in \cref{tab:Example2FEAExpANValues}. The corresponding  stretch ratios, $\mu$, and stiffness values, $K$, calculated using these results are given in \cref{tab:Example2FEAExpANSRAndStiffness}. A comparison of the analytical values with FEM and experimental results show a maximum error of less than $5\%$, (see \cref{tab:ComparisonExample2FEAExpANSRAndStiffness}), indicating a good agreement between the analytical model, numerical simulation, and experimental validation. 
\begin{table*}[h]
\centering
\renewcommand{\arraystretch}{1.5}
\resizebox{\textwidth}{!}{\begin{tabular}{|c|l|l|l|l|l|l|l|l|l|}
\hline
\text{Ex. No} & $\mu$ & \begin{tabular}{@{}c@{}}\%Error $\Delta_X$\\  (An vs FEM)\end{tabular} & \begin{tabular}{@{}c@{}}\%Error $\Delta_Y$\\  (An vs FEM)\end{tabular} & \begin{tabular}{@{}c@{}}\%Error $\Delta_X$\\  (An vs Exp)\end{tabular} &\begin{tabular}{@{}c@{}}\%Error $\Delta_Y$\\ ( An vs Exp)\end{tabular}& \begin{tabular}{@{}c@{}}\%Error $\mu$\\  (An vs FEM)\end{tabular} &\begin{tabular}{@{}c@{}}\%Error $\mu$\\ (An vs Exp)\end{tabular}&\begin{tabular}{@{}c@{}}\%Error $K$\\  (An vs FEM)\end{tabular}&\begin{tabular}{@{}c@{}}\%Error $K$\\ (An vs Exp)\end{tabular} \\
\hline
\multirow{2}{*}{2} & $\mu_p$ & $1.4739$ & $0.0615$ & $0.0300$ & $2.7077$ & $1.4132$ & $2.8140$ & $1.4960$&$0.0300$ \\ \cline{2-10}
  & $\mu_n$ & $0.0068$ & $0.01400$ & $0.0068$ & $3.5569$ & $0.0072$ & $3.4414$& $0.0068$&$0.0068$ \\
\hline
\end{tabular}}
\caption{Comparison of FEM, Analytical(An), and Experimental(Exp) Results for Example 2}
\label{tab:ComparisonExample2FEAExpANSRAndStiffness}
\end{table*}

The next section describes the design of EDCM and presents an illustrative example demonstrating the stretch ratio design with EDCM-integrated links.
%%%%%%%%%%%%%%%%%%%%%%%%%%%%%%%%%%%%%%%%%%%%%
\section{Mechanism with the EDCM}
As mentioned before, the  EDCM \label{sec:EDCM}\cite{mehul2023} functions as bistable element that can seamlessly connect and disconnect any link into which it is integrated. Thus, it facilitates switching the Poisson's effect of the mechanism by controlling connectivity in interconnecting links and enables engagement or disengagement of the MO part from the MI part. In this section, we validate the EDCM design by focusing on its role in switching the Poisson’s effect.

The EDCM is integrated into the $\Lambda$ part of the mechanism. When  engaged  (see Fig. \ref{fig:Experimental setup} A), it stiffens the link, resulting in a negative stretch ratio response corresponding to $\mu_N$. Conversely, when disengaged (Fig. \ref{fig:Experimental setup}B), the link offers zero stiffness, and the mechanism exhibits a positive stretch ratio response,  $\mu_P$.

\subsection{Example 3: $\mu_N=-1$, $\mu_P=3.87$
}

\subsubsection{Design of Mechanism}
To fabricate the prototype with the EDCM integrated to the mechanism, we consider the model with geometrical parameters corresponding to Example 3  given in  \cref{tab: Geometrical parameters}. The parameters were chosen considering the ease and feasibility of fabrication using 3D printing. The corresponding displacements and stretch ratios are computed using \cref{eq:XDispCombined,eq:YDispCombined,eq:SRP,eq:SRN}, yielding $\mu_N=-1$ and $\mu_P=3.87$.

\subsubsection{Design of EDCM}
\begin{figure*}[!htbp]
    \centering
    \includegraphics[width=0.7\linewidth]{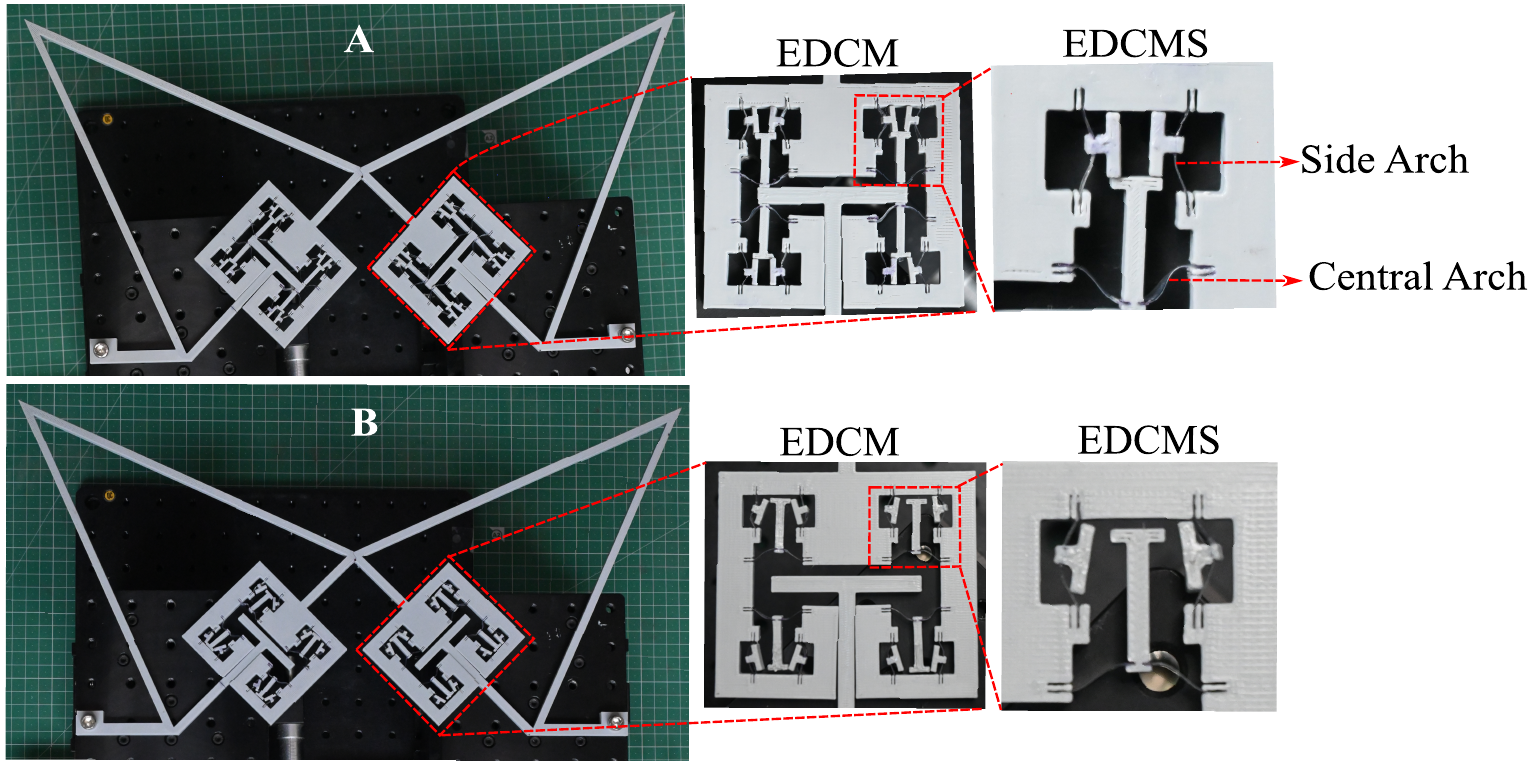}
    \caption{Experimental setup for Example 3 with EDCM in $\Lambda$ part, (A) - EDCM in engaged state(EDCM and EDCMS in the enlarged view), (B) - EDCM in disengaged state }
    \label{fig:Experimental setup}
\end{figure*}

The EDCM is composed of four switches (EDCMS) (see \cref{fig:Experimental setup}), each consisting of bistable side and central arches  \cite{mehul2023}. The EDCMS considered here has a central arch with a span of 21.5 mm and a thickness of 0.15 mm, along with two sides having a span of 16 mm and a thickness of 0.3 mm.   

\subsection{Validation of Example 3}
\begin{table*}[]
\centering
\resizebox{\textwidth}{!}{\begin{tabular}{|c|l|l|l|l|l|l|l|l|l|l|l|}
\hline
\multicolumn{1}{|l|}{Ex.} & State & Force (N) & \begin{tabular}{@{}c@{}}$\Delta_X$ ($\times10^{-3}$m) \\ (FEM)\end{tabular} & \begin{tabular}{@{}c@{}}$\Delta_Y$ ($\times10^{-3}$m)\\ (FEM)\end{tabular}& \begin{tabular}{@{}c@{}}$\Delta_X$ ($\times10^{-3}$m)\\ (Analytical)\end{tabular}& \begin{tabular}{@{}c@{}}$\Delta_Y$ ($\times10^{-3}$m)\\ (Analytical)\end{tabular} & \begin{tabular}{@{}c@{}}$\Delta_X$ ($\times10^{-3}$m)\\ (Experiment)\end{tabular} & \begin{tabular}{@{}c@{}}$\Delta_Y$ ($\times10^{-3}$m)\\ (Experiment)\end{tabular} & \begin{tabular}{@{}c@{}} $\mu$\\ (FEM)\end{tabular} & \begin{tabular}{@{}c@{}}$\mu$ \\ (Analytical)\end{tabular} & \begin{tabular}{@{}c@{}}$\mu$ \\ (Experiment)\end{tabular} \\ \hline
\multirow{2}{*}{3} & $\mu_P$ & 0.6505 & $9.4340$ & $1.2194$ &  $9.4202$ &  $1.2168$ &  $9.4340$ &  $1.2720$ & 3.8683 & 3.8709 & 3.7083 \\ \cline{2-12} 
 & $\mu_N$ & 1.8416 & $2.6190$ & $-1.3062$ & $2.6099$ &$-1.3041$& $2.6190$ & $-1.3140$& -1.0025 & -1.0007 & -0.9966 \\ \hline
\end{tabular}}
\caption{Results of FEM, Analytical, Experiment for Example 3}
\label{tab: Exp FEA values}
\end{table*}
\begin{table*}[]
\centering
\resizebox{\textwidth}{!}{\begin{tabular}{|c|c|c|c|c|c|c|c|}
\hline
Ex. No &  State &  \begin{tabular}{@{}c@{}}\%Error $\Delta_X$\\Analytical vs FEM\end{tabular}  & \begin{tabular}{@{}c@{}}\%Error $\Delta_Y$\\Analytical vs FEM\end{tabular}   & \begin{tabular}{@{}c@{}}\%Error $\mu$\\Analytical vs FEM\end{tabular}  &\begin{tabular}{@{}c@{}}\%Error $\Delta_X$\\Analytical vs Experiment\end{tabular} &\begin{tabular}{@{}c@{}}\%Error $\Delta_Y$\\Analytical vs Experiment \end{tabular}  & \begin{tabular}{@{}c@{}}\%Error $\mu$\\Analytical vs Experiment\end{tabular} \\ \hline
\multirow{2}{*}{3} &  $\mu_P$ & 0.1465 & 0.2137 & 0.0670 & 0.1465 & 4.5365 & 4.1995\\ \cline{2-8} 
 &  $\mu_N$ & 0.3487 & 0.1610 & 0.1873 & 0.3487 & 0.7591
 & 0.4074\\ \hline
\end{tabular}}
\caption{Comparison of Analytical, FEA, Experiment for Example 3}
\label{tab: ComparisonFEAAnExp}
\end{table*}The displacements of the points E and N are extracted from the experiments, FEA, and analytical model
are tabulated in  \cref{tab: Exp FEA values}.
As evident from the comparative analysis presented in \cref{tab: ComparisonFEAAnExp}, the percentage error consistently remains below $5\%$ across all measurements, indicating good agreement in our results.

\section{Biological Aspect} \label{sec:BiologicalAspect}

To grow cells on the mechanism, the mechanism must be fabricated at the microscale. In Example 4, we assess whether our model yields feasible results at this scale.

Fabrication is planned using the two-layer lithography process we previously reported \cite{Marwah2023,Marwah2024}. Based on the resolution and reliability of the lithography and the field of view of the microscope available to us, the minimum beam width was determined to be 3 $\mu$~m and the footprint of the mechanism was determined to be within a 1 cm square respectively.

\begin{figure}[!htpb]
    \centering
    \includegraphics[width=0.6\linewidth]{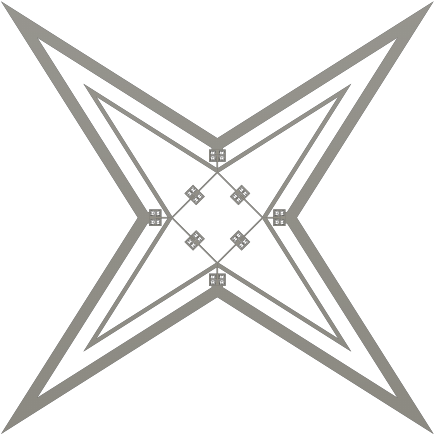}
    \caption{Model for Microfabrication with equalized stiffness of 1 N/m in both positive and negative stretch states.}
    \label{fig:MicrofabStiffness}
\end{figure}

The out-of-plane thickness $h$ for both the M and $\Lambda$ parts is set to 5 $\mu$~m , with the minimum in-plane dimension corresponding to the thickness of the bistable arch at 3$\mu$~m . A design satisfying  $\mu_P = 1.5$, $\mu_N = -1$, and a stiffness of 1 N/m in both stretch states was designed (see \cref{fig:MicrofabStiffness}), with dimensions listed in \cref{tab:Geometrical parameters of model for microfabrication}. The final model has a footprint under 8 mm square.

\begin{table*}[h]
\setlength{\tabcolsep}{0.4\tabcolsep}
\centering
\begin{tabular}{|c|c|c|c|c|c|c|c|}
\hline
\begin{tabular}{@{}c@{}}Ex. \\No. \end{tabular}  & \begin{tabular}{@{}c@{}}$a$ \\(mm) \end{tabular} & \begin{tabular}{@{}c@{}}$a_{\text{O}}$ \\(mm) \end{tabular} & \begin{tabular}{@{}c@{}}$\theta_1$ \\($^o$) \end{tabular} & \begin{tabular}{@{}c@{}}$h$ \\($\times10^{-3}$mm) \end{tabular} & \begin{tabular}{@{}c@{}}$t_{\text{M}}$ \\($\times10^{-3}$mm) \end{tabular}    & \begin{tabular}{@{}c@{}}$t_{\Lambda}$ \\($\times10^{-3}$mm) \end{tabular}    & \begin{tabular}{@{}c@{}}$t_{\text{O}}$ \\($\times10^{-3}$mm) \end{tabular}    \\ \hline
4       & 0.85  & 1.322  & 57.69  & 5     & 72.928 & 24.235 & 504.204 \\ \hline
\end{tabular}
\caption{Geometrical Parameters of models for microfabrication}
    \label{tab:Geometrical parameters of model for microfabrication}
\end{table*}

\section{Summary and future work} \label{sec:SummaryAndFutureWork}
This paper presents the design of a compliant mechanism with invertible and tunable Poisson’s ratio and stiffness. Inversion of the Poisson’s ratio (or stretch ratio) and stiffness tunability are achieved by incorporating bistable EDCMs into the inner links of the mechanism. The stretch behavior and stiffness are tuned by selecting appropriate geometrical parameters of the MI and MO parts, as captured by an analytical model.  We aim to utilize the mechanism to investigate how substrate mechanical properties influence the behavior of biological cells. 

The analytical model relating stretch ratio and stiffness to the mechanism’s geometry is validated through FEA and tabletop experiments using 3D-printed prototypes across illustrative example cases. A microscale version of the design has also been developed (\cref{fig:MicrofabStiffness}). As the next step, we will microfabricate the mechanism, culture cells on it, and investigate how inverting the Poisson’s ratio influences cellular behavior and function.

\label{sec:summary}

\section*{Acknowledgments}

We gratefully acknowledge the funding received from the
 Department of Science and Technology, Government of India
 (SRG/2021/001463) and IIT-H SG-92.
\vspace{6 mm}
\bibliographystyle{plain}
\bibliography{references}

\vskip2pc

\end{document}